\documentclass[a4paper,fleqn,useAMS,usenatbib]{mnras}

\usepackage{graphicx}	% Including figure files
\usepackage{amsmath}	% Advanced maths commands
\usepackage{amssymb}	% Extra maths symbols
\usepackage{multicol}        % Multi-column entries in tables
\usepackage{bm}		% Bold maths symbols, including upright Greek
\usepackage{pdflscape}	% Landscape pages
\usepackage{xspace}
\usepackage[final]{pdfpages}
\usepackage{layouts}

\usepackage{ulem}

\newcommand{\gv}{$\gamma^2$~Vel\xspace}
\newcommand{\brite}{{\it BRITE-Constellation}\xspace}

\title[Orbital variability of the \textit{BRITE}st Wolf-Rayet star]{The variability of the \textit{BRITE}-est Wolf-Rayet binary, \\ $\gamma^2$ Velorum I. Photometric and spectroscopic evidence for colliding winds }

\author[Richardson et al.]{Noel D.~Richardson$^1$\thanks{E-mail:noel.richardson@utoledo.edu}
Christopher M.\ P.\ Russell$^{2}$,
Lucas St-Jean$^{3}$,
Anthony F. J. \newauthor Moffat$^{3}$,
Nicole St-Louis$^{3}$,
Tomer Shenar$^{4}$,
Herbert Pablo$^{3}$,
Grant M. Hill$^{5}$,
\newauthor Tahina Ramiaramanantsoa$^{3}$,
Michael Corcoran$^{2,6}$,
Kenji Hamuguchi$^{2}$,
\newauthor Thomas Eversberg$^{7}$,
Brent Miszalski$^{8, 9}$,
Andr\'e-Nicolas Chen\'e$^{10}$,
Wayne Waldron$^{11}$,
\newauthor Enrico J. Kotze$^{8}$,
Marissa M. Kotze$^{8, 9}$,
Paul Luckas$^{12}$,
Paulo Cacella$^{13}$,
\newauthor Bernard Heathcote$^{13}$,
Jonathan Powles$^{13}$,
Terry Bohlsen$^{13}$,
Malcolm Locke$^{13}$,
\newauthor Gerald Handler$^{14}$,
Rainer Kuschnig$^{15, 16}$,
Andrzej Pigulski$^{17}$
Adam Popowicz$^{18}$,
\newauthor Gregg A. Wade$^{19}$, 
and Werner W. Weiss$^{15}$\\
$^{1}$ Ritter Observatory, Department of Physics and Astronomy, The University of Toledo, Toledo, OH 43606-3390, USA\\
$^{2}$ X-ray Astrophysics Laboratory, NASA Goddard Space Flight Center, Greenbelt, MD 20771, USA\\
$^{3}$ D\'epartement de physique and Centre de Recherche en Astrophysique du Qu\'ebec (CRAQ), Universit\'e de Montr\'eal, C.P. 6128,\\  Succ.~Centre-Ville, Montr\'eal, Qu\'ebec, H3C 3J7, Canada\\
$^{4}$ Institut f\"{u}r Physik und Astronomie, Universit\"{a}t Potsdam, Karl-Liebknecht-Str. 24/25, D-14476 Potsdam, Germany\\
$^{5}$ W.M. Keck Observatory, 65-1120 Mamalahoa Hwy, Kamuela HI, 96743\\
$^{6}$ Institute for Astrophysics and Computational Sciences, Department of Physics, The Catholic University of America, Washington, DC 20064, USA \\
$^{7}$ Schn\"orringen Telescope Science Institute, Waldbr\"ol, Germany\\
$^{8}$ South African Astronomical Observatory, PO Box 9, Observatory, 7935, South Africa\\
$^{9}$ Southern African Large Telescope Foundation, PO Box 9, Observatory, 7935, South Africa\\
$^{10}$ Gemini Observatory, Northern Operations Centre, 670 North A'ohoku Place, Hilo, HI, 96720, USA \\
$^{11}$ Eureka Scientific, Inc., 2452 Delmer Street, Oakland, CA 94602, USA \\
$^{12}$ International Centre for Radio Astronomy Research, University of Western Australia, 35 Stirling Hwy, Crawley, WA 6009, Australia \\
$^{13}$ SASER \\
$^{14}$ Centrum Astronomiczne im. M. Kopernika, Polska Akademia Nauk, Bartycka 18, 00-716 Warszawa, Poland \\
$^{15}$ Institut f\"{u}r Astrophysik, Universit\"{a}t Wien, T\"{u}rkenschanzstrasse 17, 1180 Wien, Austria \\
$^{16}$ Institut f\"{u}r Kommunnikationsnetze und Satellitenkommunikation, Technical University Graz, Infeldgasse 12, 8010 Graz, Austria \\
$^{17}$ Instytut Astronomiczny, Uniwersytet Wroc?awski, Kopernika 11, 51-622, Wroc?aw, Poland \\
$^{18}$ Institute of Automatic Control, Silesian University of Technology,, Akademicka 16, 44-100 Gliwice, Poland \\
$^{19}$ Department of Physics, Royal Military College of Canada, PO Box 17000 Kingston, ON K7K 7B4, Canada 
}

\begin{document}
%%\received{}
%%\accepted{}
%\bibliographystyle{mnras}

\date{}

\pagerange{\pageref{firstpage}--\pageref{lastpage}} \pubyear{2017}

\maketitle

\label{firstpage}

\begin{abstract}

We report on the first multi-color precision light curve of the bright Wolf-Rayet binary $\gamma^2$ Velorum, obtained over six months with the nanosatellites in the {\it BRITE-Constellation} fleet. In parallel, we obtained 488 high-resolution optical spectra of the system. In this first report on the datasets, we revise the spectroscopic orbit and report on the bulk properties of the colliding winds. We find a dependence of both the light curve and excess emission properties that scales with the inverse of the binary separation. When analyzing the spectroscopic properties in combination with the photometry, we find that the phase dependence is caused only by excess emission in the lines, and not from a changing continuum. We also detect a narrow, high-velocity absorption component from the He I $\lambda$5876 transition, which appears twice in the orbit. We calculate smoothed-particle hydrodynamical simulations of the colliding winds and can accurately associate the absorption from He I to the leading and trailing arms of the wind shock cone passing tangentially through our line of sight. The simulations also explain the general strength and kinematics of the emission excess observed in wind lines such as C~III $\lambda 5696$ of the system. These results represent the first in a series of investigations into the winds and properties of $\gamma^2$ Velorum through multi-technique and multi-wavelength observational campaigns.

\end{abstract}


\section{Introduction}

%\printinunitsof{in}\prntlen{\textwidth} \printinunitsof{cm}\prntlen{\textwidth}
%
%\printinunitsof{in}\prntlen{\paperwidth} \printinunitsof{cm}\prntlen{\paperwidth}
%
%
%\printinunitsof{in}\prntlen{\textheight} \printinunitsof{cm}\prntlen{\textheight}
%
%\printinunitsof{in}\prntlen{\paperheight} \printinunitsof{cm}\prntlen{\paperheight}
%
%
%\printinunitsof{in}\prntlen{\oddsidemargin} \printinunitsof{cm}\prntlen{\oddsidemargin}
%
%\printinunitsof{in}\prntlen{\evensidemargin} \printinunitsof{cm}\prntlen{\evensidemargin}
%
%\printinunitsof{in}\prntlen{\topmargin} \printinunitsof{cm}\prntlen{\topmargin}
%
%
%\printinunitsof{in}\prntlen{\hoffset} \printinunitsof{cm}\prntlen{\hoffset}
%
%\printinunitsof{in}\prntlen{\voffset} \printinunitsof{cm}\prntlen{\voffset}
%
%
%\printinunitsof{in}\prntlen{\leftmargin} \printinunitsof{cm}\prntlen{\leftmargin}
%
%\printinunitsof{in}\prntlen{\rightmargin} \printinunitsof{cm}\prntlen{\rightmargin}

Massive stars provide important feedback to their parent galaxies through ionizing radiation, stellar winds, and their terminal supernova explosions. A growing amount of evidence suggests that they virtually all start their lives in multiple systems, and interactions between binary components is common \citep{2012Sci...337..444S}. During the course of binary evolution, a majority of interactions involve matter transferring from one star to the other. Sometimes, they merge into a single rapidly rotating star \citep{2012ASPC..465..342V,2012A&A...538A..75P}, but the majority of the time they undergo Roche lobe overflow (RLOF) from the primary to spin up the companion star \citep[e.g.,][]{2017MNRAS.464.2066S}. Through either the RLOF process or mass lost through its strong stellar wind, the primary star can lose its envelope and become a classical He-burning Wolf-Rayet (WR) star.

WR stars are characterized by strong stellar winds and high effective temperatures. They represent the final stage of a massive star's life as it evolves through the helium main sequence. The mass loss rates are high ($\dot{M} \sim 10^{-5} M_\odot {\rm yr}^{-1}$) as are the terminal wind speeds ($v_\infty \sim 1500-3000$ km s$^{-1}$). Many WR stars are in binary systems, and the companion stars tend to be massive O stars \citep[e.g.,][]{2015MNRAS.447.2322R}. The winds collide and produce a shock cone, even in relatively long-period binaries \citep[e.g.,][]{2001MNRAS.324...33B}. Orbital modulation can be observed with spectroscopy, but often with photometry as well, even in non-eclipsing systems \citep[e.g.,][]{1996AJ....112.2227L}. However, the fact that WR stars are near the end of their lives makes the measurement of their masses difficult due to their scarcity, and only four systems have been resolved with interferometry where visual orbits are attainable in a reasonable amount of time: \gv \citep{North, 1970MNRAS.148..103H, 2007A&A...464..107M}, WR 140 \citep{2004ApJ...602L..57M, 2011ApJ...742L...1M}, and WR 137 and WR 138 \citep{2016MNRAS.461.4115R}. Some other long-period binaries were resolved with the {\it Hubble Space Telescope} \citep[e.g., ][]{1998AJ....115.2047N, 2002ASPC..260..407W, 2005AJ....130..126W}, but these systems will have much longer periods, making it difficult to obtain both visual and spectroscopic orbits. However, only \gv and WR 140 have enough time-coverage with interferometry to constrain the masses of the component stars using their visual orbits. However, in these systems, phase coverage to constrain phase-resolved variability is difficult to obtain due to the longer periods of the systems, but some efforts have been successful, especially for WR 140 near periastron \citep{2003ApJ...596.1295M, 2011MNRAS.418....2F}.

There are a few WR stars that were observed with precision time-series photometry with the {\it MOST} ({\it Microvariability
and Oscillations of STars}) space telescope \citep{2003PASP..115.1023W}. These data are complicated in that they show some quasi-periodic variations which may be related to large-scale structures in the winds of these stars \citep[e.g., WR 110; ][]{2011ApJ...735...34C}. In the case of WR 123, \citet{2005ApJ...634L.109L} found that the star may pulsate with a period of 10 h, but such a convincing pulsation has not been observed for any other WR. Several WR stars show evidence of a photometric eclipse-like phenomenon, where electron scattering causes symmetric dips in their light curves \citep{1996AJ....112.2227L}. Atmospheric eclipses were seen in the WR binary CV Serpentis, observed with the {\it MOST} satellite by \citet{2012MNRAS.426.1720D}, but the eclipses changed with each cycle. The authors attributed the changing eclipse depths as evidence for sporadic dust formation within the wind of the WR star. However, these WR stars observed with {\it MOST} tend to be relatively faint, so simultaneous spectroscopic variability campaigns were difficult to coordinate due to their faintness and telescope oversubscription rates. This study aims to rectify these limitations to our understanding of the variability.

\gv is the closest and brightest WR star at a distance of only 330$^{+8}_{-7}$ pc \citep{North}. It is a double-lined spectroscopic binary, composed of a WC8 star and an O7.5 giant star, with an orbital period of $78.53\pm0.01$ d \citep{1997A&A...328..219S}. The individual component spectra were modeled with the non-LTE radiative transfer code CMFGEN by \citet{1999A&A...345..163D} for the O star and \citet{deMarcoWR} for the WR star. The spectroscopic orbit is well constrained and was combined with an interferometric analysis of the orbit by \citet{North}. The combination of these studies led to very precise orbital parameters such as the inclination of $65.5\pm0.4^\circ$ and stellar masses ($M_{\rm WR} = 9.0 \pm 0.6 M_\odot$; $M_{\rm O} = 28.5 \pm 1.1 M_\odot$), allowing us to study the binary in detail and constrain models. The system is likely the byproduct of binary interactions \citep{eldridge}.

The stellar properties from interferometry were explored by \citet{2007A&A...464..107M}. They used the spectral models of \citet{1999A&A...345..163D} and \citet{deMarcoWR} to model spectral observations from the VLTI/AMBER instrument. Their findings supported a distance that was larger than that measured by the {\it Hipparcos} mission, a result that was then confirmed by \citet{North}. They showed that a slight elongation of the WR wind was compatible with the observed visibilities from the emission lines. Unfortunately, the only interferometric observation with spectral dispersion for \gv was this single point from VLTI obtained during the commissioning phase of the instrument. 

\citet{2017arXiv170101124L} examined new VLTI/AMBER data and used numerical simulations to derive an orbit that was nearly identical to that reported by \citet{North}. Their numerical simulations allowed them to discern the effect of the spatially-resolved colliding winds' shock cone in the near-infrared continuum and in emission lines. Further, they found phase-dependent behavior of the P Cygni absorption component of the He I $\lambda$2.059 $\mu$m transition, likely originating from the shock cone as it orbits. The emission lines in the $K$-band show excess strength near periastron, but the temporal coverage of their interferometric observations does not allow much information about the orbital dependence of the variations. 

\gv is the closest example of a colliding-winds (CW) binary in the Galaxy. In CW systems with WR components, we typically can observe evidence of the shock cone in the optical spectral lines \citep[e.g., ][]{2000MNRAS.318..402H}, X-ray production \citep[e.g., ][]{2015PASJ...67..121S}, and in some WC stars dust is produced \citep[e.g., ][]{1990MNRAS.243..662W}. It is currently unknown if all the later-type WC stars require a companion wind to produce dust or not. However, some earlier type WC stars (e.g., WR 140) produce dust through their wind collisions. The high-energy photons come from the conversion of kinetic energy to radiation in the shock cone, and in the case of \gv has been shown to produce $\gamma$-rays \citep{2016MNRAS.457L..99P}. \gv is a well-studied X-ray source, with variability studies previously accomplished related to its {\it ROSAT} light curve \citep{1995A&A...298..549W}, while {\it ASCA} and {\it XMM-Newton} have been used to examine the spectral differences between the high and low states, which shows an extremely variable hard X-ray component \citep{2000MNRAS.316..129R, 2004A&A...422..177S}. The binary has a small enough orbit that radiative braking likely plays a large role in the X-ray production \citep{1997ApJ...475..786G}.

The \brite nanosatellites observed \gv photometrically for six contiguous months during 2014--2015. These measurements represent the best photometric time-series of any WR star to date, with milli-mag precision. In parallel, we scheduled and executed a world-wide spectroscopic campaign on the system. This paper presents the first analysis of these data, concentrating on the optical variability related to the orbit of the system. Future analyses will examine short-term variations, detailed modeling of the colliding winds, and recent X-ray observations with {\it XMM-Newton}. The remainder of this paper is structured as follows. In Section 2, we present the observations. Section 3 details the properties of the spectroscopic orbit we derived. The photometric results are analyzed in Section 4, while the spectroscopic variations from the colliding winds are discussed in Section 5. We discuss the results in Section 6 and outline the future parts of an effort to understand the nearest colliding winds binary in Section 7 along with our main conclusions.

\section{Observations}

\subsection{BRITE Photometry}

Precision time-series photometry was collected for this project with {\it BRITE-Constellation} ({\it BRIght Target Explorer}). {\it BRITE-Constellation} is a fleet of five nanosatellites described in detail by \citet{2014PASP..126..573W} and \citet{2016PASP..128l5001P}. The telescopes have 3 cm apertures, with two satellites recording blue images (3900--4600\AA) and three satellites recording red images (5500--7000\AA). Their field of view is large, $20^\circ \times24^\circ$. Data are downloaded for up to 20--30 stars simultaneously, depending on the field. The orbital periods of the fleet range between 97 and 101 minutes. To allow for performing reliable photometry, the optical system has intentional defocussing applied. The pixel scale is 28\arcsec, and the FWHM for stellar PSFs are between 5--8 pixels, meaning that the light of $\gamma^1$ Vel and the other surrounding stars blurs into the light of \gv during the observations. However, their contribution is small ($\sim 10\%$) in comparison to the brightness of \gv\, so the variability observed is strongly representative of the WC$+$O system.

The observations were recorded between 2014 December 1 and 2015 May 28, and a total of 116,692 data points were recorded with the {\it BRITE-Austria} (blue filter) and {\it BRITE- Toronto} (red filter) satellites. The measurements were made with continuous one-second exposures for 5--10 minutes per orbit when the field was visible. Further reductions were done to remove artifacts introduced by an extraordinary and unstable dark current rate in so-called hot pixels and in warm columns. Precision was increased by creating a mean measurement for each satellite orbit of 100.36 minutes ({\it BRITE-Austria}) or 98.24 minutes ({\it BRITE-Toronto}), resulting in a $\sim 1$mmag precision for the differential photometry after removing the correlations with instrumental effects, in the manner described by \citet{2016A&A...588A..55P}.

We also note that the \brite nanosatellites were switched from operating in a stare imaging mode to a chopping mode during the course of these observations. The chopping mode performs photometry through a difference between consecutive frames, in which the telescope pointing is offset, resulting in the stellar PSF shifting. The difference images obtained by subtracting two consecutive frames, have virtually all the defects subtracted with the positive and negative PSFs profile clearly visible. This process has been shown to increase data quality and precision \citep{2016SPIE.9904E..1RP, 2017arXiv170509712P}.

\subsection{Ground-based Spectroscopy}

We initiated a world-wide campaign to better determine the orbit and spectroscopic variability of \gv during the course of the \brite campaign. We summarize the sources of spectroscopic data in Table 1, and summarize the reductions in the following subsections. We concentrated on the C~III $\lambda 5696$ transition for investigation of the variability due to the atomic processes that contribute to its formation. \citet{2015wrs..conf...65H} describe the emission of this line ($2s3d ^1D \rightarrow 2s2p ^1P^0$), and how the population in the $2s3p ^1P^0$ state is drained by transitions to $2p^2\ ^1D$ and $2p^2\ ^1S$, allowing this line to remain optically thin and flat-topped despite strong emission. C~III $\lambda$5696 is sensitive to density effects and is also free from blended wind lines, making it ideal to study colliding winds and clumping processes in the WR wind. We also investigated other lines when possible or appropriate, as described in subsequent sections.

\begin{table*}
\centering
\begin{minipage}{180mm}
\caption{Spectroscopic Observing Log \label{table1}}
\begin{tabular}{lllllccll}
\hline
Location/Observer & Telescope               &       Spectrograph         & CCD                           &  Range (\AA)                  &       $R$             &       $N_{\rm spectra}$       &       HJD (first)             &       HJD (last)      \\
\hline
\multicolumn{9}{c}{{\it Professional Facilities}} \\ \hline
CTIO		& 	1.5 m      &       CHIRON echelle        &                          &       4500--8500                      &       80,000  &                 216                   &       2457025             &       2457199     \\
SAAO 		&	1.9 m      &       GIRAFFE echelle       &                          &       4225--7000                      &       39,200  &                  29                   &       2457051             &       2457064     \\ \hline
\multicolumn{9}{c}{{\it Amateur Observers}} \\ \hline
Heathcote              & 0.28 m C &   LHIRES III      &      Atik314L               &                   5640--5780              &    9456                   &            43                  &  2457041       &  2457094      \\
Luckas          &    0.36 m RC    &   LHIRES III          &      Atik314L               &                   5630--5770              &    9351                   &           121                  &  2456971       &  2457195      \\
Cacella         &    0.5  m       &        LHIRES III                 &         ST8XME           &                   5590--5790              &    7609                   &            56                  &  2457041       &  2457110      \\
Powles          &    0.25 m SCT   &   Spectra L200        &      Atik 383L              &                   5540--5950              &    8549                   &            19                  &  2457010       &  2457147      \\
Bohlsen         &    0.28 m C     &   Spectra L200        &      ST8XME                  &                   5480--5875              &    7101                   &            12                  &  2457051       &  2457120      \\
%Locke           &                 &                       &                              &                                           &                           &                                &                &               \\
\hline
\end{tabular}
\end{minipage}
\end{table*}

\subsubsection{CTIO 1.5 m/CHIRON}

We began monitoring \gv with the CTIO 1.5m telescope operated by the SMARTS Consortium and the high-resolution CHIRON echelle spectrograph \citep{2013PASP..125.1336T} that covers the optical spectrum from $\sim4500$\AA\ to $\sim8500$\AA. The starlight is fed through a multi-mode fibre that has a size of 2.7\arcsec on the sky. We began monitoring the system with the ``slicer" mode ($R \sim 80,000$) in 2015 February, which we continued until the star was too low in the sky to continue observations (2015 June). On a given night, we obtained between 1 and 6 spectra, each of which consisted of a co-added set of three 20-s exposures, which combined to give a S/N of $\gtrsim 400$ per pixel in most observations.
All spectra were corrected for bias and flat field effects and wavelength calibrated through the standard CHIRON pipeline. This leaves a very strong blaze function present on each order which is difficult to remove in the presence of strong emission lines, frequently exhibiting widths similar to the echelle order range. Therefore, we observed HR 4468 (B9.5Vn) to fit the continuum blaze function empirically on orders without spectral lines as was done by \citet{2016MNRAS.461.2540R} for $\eta$ Carinae. Echelle orders with strong spectral lines present (e.g., around H$\alpha$ and H$\beta$) had blaze functions interpolated between the adjacent orders. The resulting spectra were then normalized and combined into standard one-dimensional spectra, and the resulting order overlaps in the blue showed us that the blaze removal was accurate to within $0.5\%$. A global normalization was then applied to obtain a continuum of unity, which we adjusted in the regions adjacent to our spectral lines of interest for analysis.

\subsubsection{SAAO}

We used the {Grating Instrument for Radiation Analysis with a Fibre
Fed Echelle (GIRAFFE)} spectrograph at the $1.9$~m telescope of the
South-African Astronomical Observatory (SAAO) to obtain $206$
high-resolution ($R\sim39,000$) echelle spectra of \gv
covering the wavelength range $4250-6800$\AA, during $14$ nights between 2015 January 28 and 2015 February 10. Camera flat-fields were acquired
at the beginning of the run whereas fiber flat-fields and bias frames were
obtained in the usual way at the beginning of each night. Arc exposures
were acquired between each set of $4-10$ consecutive spectra of \gv and
after switching back from observations of the other scientific target of
the campaign ($\zeta$ Puppis; Ramiaramanantsoa et al., in prep). Data reduction and extraction of
one-dimensional wavelength-calibrated spectra uncorrected for the blaze
function were performed with {\tt indlulamithi}, the standard
Python-based pipeline for {GIRAFFE} data reduction and extraction.
Correction for the blaze function in the extracted spectra of
\gv was achieved by dividing the spectra by those of the B9.5V
star HR 4468 taken with the same instrument setup, and in a similar manner to the CHIRON data. After stacking consecutive exposures, we obtained $29$ spectra having a S/N$\gtrsim 100$ in the
continuum.

\subsubsection{SASER}

The advent of affordable spectrographs available for small telescopes has allowed for a new era in massive star research, especially in the case of WR stars \citep[e.g.,][]{2011MNRAS.418....2F,2016MNRAS.460.3407A}, and has led to a Southern Astro Spectroscopy Email Ring\footnote{http://saser.wholemeal.co.nz/} (SASER) which promotes professional/amateur collaboration, allowing for results to be shared quickly within the group, which has benefited several campaigns on massive stars \citep[e.g.,][]{2016ApJ...819..131T}. During the course of the {\it BRITE-Constellation} photometric campaign, we initiated a collaboration with these southern spectroscopists who have small telescopes (0.25--0.5 m) and commercially available spectrographs such as the LHIRES III\footnote{http://www.shelyak.com/rubrique.php?id\_rubrique=6\&lang=2} or Spectra L200\footnote{http://www.jtwastronomy.com/products/spectroscopymain.html}, which can deliver spectra with resolving power of 7,000--15,000. The details of these instruments are given in Table 1.
All 250 spectra were corrected with bias, dark, and flat field frames taken near in time to the science exposures, and were wavelength calibrated with an internal lamp taken either before or after the stellar exposures. The wavelength solution was confirmed with telluric absorption lines near the C III $\lambda 5696$ line.

\subsection{HEROS archival data}

In addition to the measurements from our data, we re-measured the data from \citet{1997A&A...328..219S} in the same manner. These data included 132 blue spectra spanning 3420--5590\AA\ and 145 red spectra spanning 5780--8632\AA. The observations were collected with the HEROS (Heidelberg Extended Range Optical Spectrograph) on the ESO 50 cm telescope. The resolving power of these observations was $\sim20,000$.

\subsubsection{Final reductions of all spectra}

The C~III $\lambda$5696 transition is contaminated by weak telluric absorption lines. We corrected all spectra for telluric absorption using a template spectrum \citep{2011ApJS..195....6W}. The high-resolution telluric spectrum was convolved to the instrumental response of the observation determined using a Gaussian fit of an isolated telluric line. We further used these lines to ensure that the wavelength solution of the spectra was accurate to within one pixel for all data. The strengths of these model telluric lines were varied in order to allow for an optimal removal of the contaminating lines. In the case of the CHIRON or GIRAFFE data, we also telluric corrected the spectrum in other regions where the absorptions were optically thin, as we did not observe a telluric standard for every observation. Finally, all spectra were normalized in a common wavelength region in order to be inter-compared.

\section{The double-lined spectroscopic orbit}

\begin{figure}
\includegraphics[angle=90,width=3.0in]{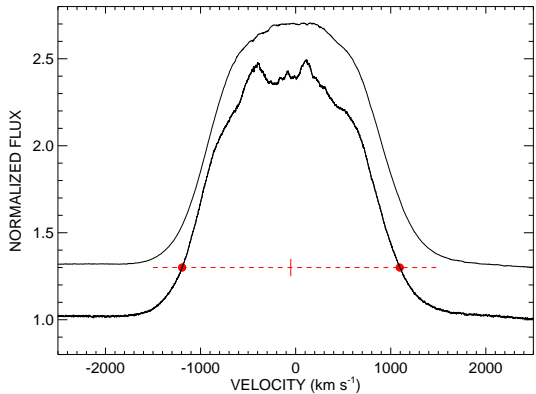}
\caption{A sample spectrum around the C III $\lambda$5696 line of \gv from the CHIRON data set is shown with the level of the bisector indicated along with the measured velocities on the line wings and the derived radial velocity. The average spectrum from the CHIRON data set is offset $+0.3$ in continuum units to illustrate the variability present in individual spectra. }
\label{WRstarvel}
\end{figure}

The purpose of this first paper on the data sets obtained on \gv is to explore the optical variability that is driven by the binary orbit. As such, we began our study with a re-derivation of the double-lined orbit. Here we outline our methodology for the measurements made and present a finalized ephemeris of the system. We began our analysis with the measurement of the radial velocities of strong WR emission lines. These lines are often contaminated by the O star absorption lines or effects of the colliding winds, so we utilized a bisector method to measure velocities near the base of the wind profiles which we illustrate in Fig.~\ref{WRstarvel}. The bisector measurements minimizes the effects of the inhomogeneities in the wind line profiles caused by the colliding winds \citep[e.g.,][]{2000MNRAS.318..402H}, co-rotating interaction regions \citep[e.g.,][]{2016MNRAS.460.3407A}, or any other short-lived variations present in the wind density.

%For the O star, we used a model spectrum with the parameters for the star given by de Marco et al.~(2001?) for temperature, $\log g$, luminosity, and rotational broadening that was calculated with the PoWR code {\bf TOMER- PLEASE ADD ANYTHING YOU THINK IS NECESSARY}. This model was then diluted in strength to match the line depths of the O star in the binary.

For the O star, we used a model spectrum with the parameters for the star given by \citet{1999A&A...345..163D} for temperature $T_{\rm eff}$, gravity $\log g_*$, luminosity $L$, and projected rotational velocity $v \sin i$. The model was calculated with the Potsdam Wolf-Rayet (PoWR) model atmosphere code \citep[e.g.,][]{2015ApJ...809..135S}. The code solves the non-LTE radiative transfer in expanding atmospheres and accounts for wind inhomogeneities and line blanketing \citep[see][for further details]{2002A&A...387..244G,2004A&A...427..697H}. The resultant model spectrum was then diluted in strength to match the line depths of the O star in the binary as the WR star contributes 38\% of the optical light \citep{North}, making the O star lines appear weaker than that of a single O star.

\begin{figure}
\includegraphics[angle=90,width=3.5in]{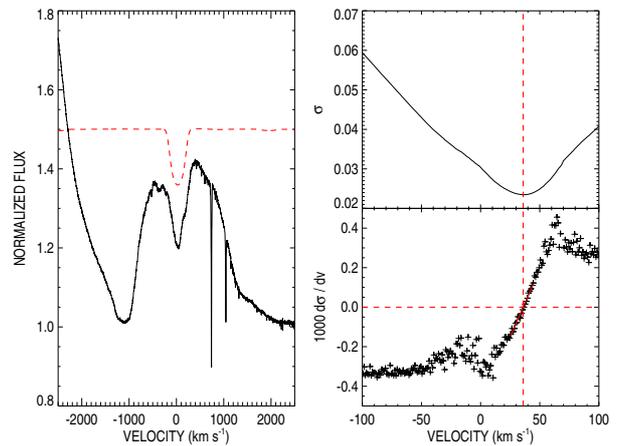}
\caption{A sample spectrum around the He I $\lambda$5876 line of \gv from the CHIRON data set is shown on the left, with the model O star spectrum also plotted, which is offset for clarity. We calculate a standard deviation of the region surrounding the absorption line when the model is subtracted for a large number of radial velocities (top right). We then determine the radial velocity by finding the zero-point of the numerical derivative (bottom right).  }
\label{Ostarvel}
\end{figure}

In Figure \ref{Ostarvel}, we show our method for measuring the velocity of the absorption lines in our spectra. We began by convolving the model to the spectral resolution of the observations which included the O star absorption lines. Then, we subtracted the model from the observations for a range of velocities larger than the amplitude of the velocity changes. As the $\gamma$-velocity is near zero, and the semi-amplitude of the O star's orbit is less than 65 km s$^{-1}$, we adopted a range of $\pm120$ km s$^{-1}$ over which to measure the velocity of the absorption line. At each test velocity, we calculated the smoothness of the WR emission peak by means of the standard deviation ($\sigma$), as every line is blended with emission from the WR star. This measurement reached a minimum at a velocity corresponding to the radial velocity of the O star during the observation. Then, we identified the radial velocity at which the numerical derivative $d\sigma / dv$ was equal to zero, and adopted this as the velocity for that observation. These measurements are shown for a sample He~I $\lambda$5876 line in Figure \ref{Ostarvel} and are available as an online supplement to this paper.

We began our analysis by checking the period of the binary orbit with time-series techniques such as the Scargle periodogram \citep{1982ApJ...263..835S} and the phase dispersion minimization routines \citep[e.g.,][]{1978ApJ...224..953S}. We found that because our data span about two orbital periods (78.53 d), we could not significantly improve upon the period's accuracy. Therefore, we adopt the 78.53 d period derived by \citet{1997A&A...328..219S}, and used in the interferometric analysis of \citet{North}.

\begin{figure*}
\includegraphics[angle=90,width=3.25in]{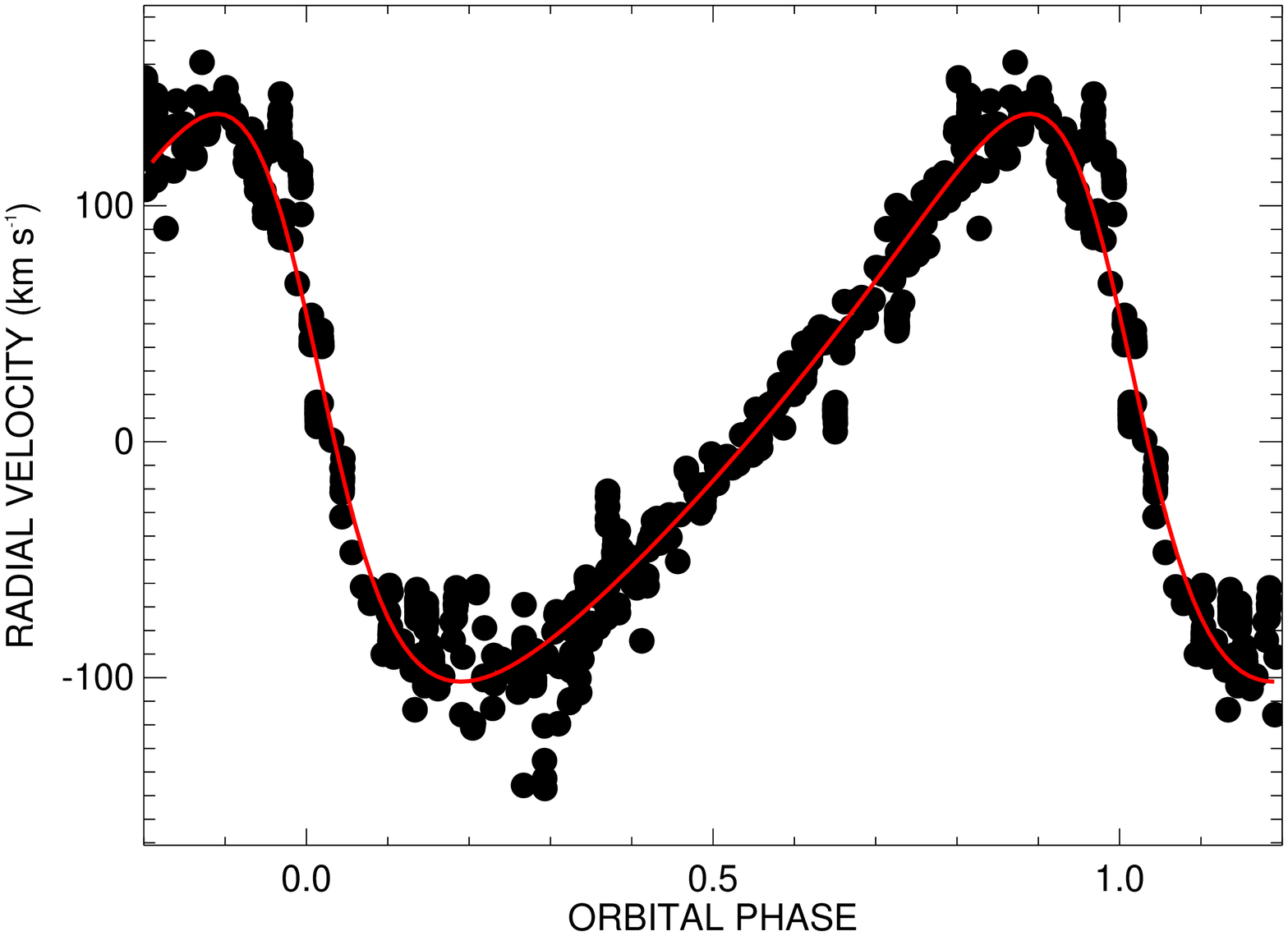}
\includegraphics[angle=90,width=3.25in]{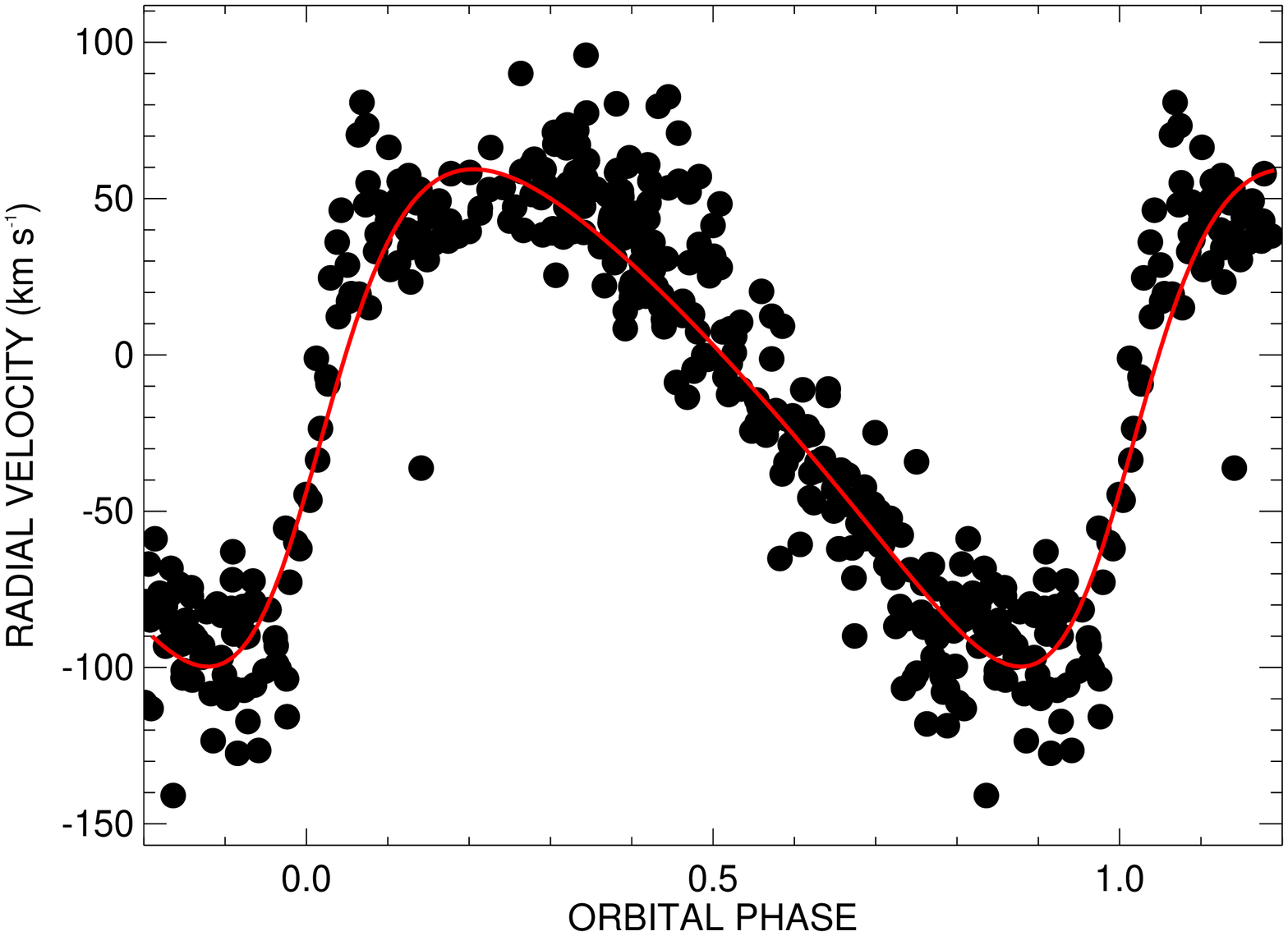}
\caption{The spectroscopic orbit \gv. On the left, we show the orbit of the WR star from the C III $\lambda$5696 measurements. The orbit of the O star from the H I + He II $\lambda$4861 measurements is shown on the right. \label{orbit} }
\end{figure*}

We fit orbital elements for each emission line's set of radial velocity measurements. Each measurement was weighted according to its spectral resolution and signal-to-noise ratio. The fit for the C III $\lambda5696$ line is shown in Fig.~\ref{orbit}, as is the O star orbit. The orbital elements for each line are tabulated in Table \ref{table-elements}. We adopt the C III $5696$ orbit as our best fit due to the similarity to the eccentricity derived in the interferometric orbit \citep{North} and the relatively small residuals in the fit. We found similar success with the O star, where the H I + He II $\lambda4861$ transition produced a good orbital fit to the data. In general, the orbital fits of \citet{North} and \citet{1997A&A...328..219S} are confirmed, and were not significantly improved upon. We adopt the spectroscopic orbital parameters of C III $\lambda5696$ for the remainder of this analysis. We anticipate that differences in the orbital elements are related to excess emission and absorption effects showing up near the periastron passage.

%\begin{landscape}
\begin{table*}
\centering
%\rotate
\caption{Orbital Elements \label{table-elements}}
\resizebox{\textwidth}{!}{
\begin{tabular}{l c c c c c c c c c}
\hline
Line	&	$\gamma$ 	&	$K$ 			&	$\omega$	&	$T_0$ 		&	$e$	&	$P$	&	$f(M)$	&	$a\sin i$	&	$N_{\rm spec}$	\\
	&	(km s$^{-1}$)	&	(km s$^{-1}$)	&	($^\circ$)	&	(HJD - 2,450,000) &		&	(d)	&	($M_\odot$) &	(AU)		&		\\ \hline
%WR star	
\multicolumn{10}{c}{WC star emission lines} \\ \hline
CIII$\lambda5696$	&	$6.9 \pm 0.5$	&	$120.4 \pm 0.7$	&	$73.0 \pm 2.2$	&	$7188.02 \pm 0.48$	&	$0.333 \pm 0.006$	&	$78.53$	&	$11.9 \pm 0.2$	&	$0.819 \pm 0.005$	&	587	\\	
CIII$\lambda6744$	&	$-175.4 \pm 0.3$	&	$115.8 \pm 0.5$	&	$72.5 \pm 0.3$	&	$7190.82 \pm 0.06$	&	$0.285 \pm 0.004$	&	$78.53$	&	$11.1 \pm 0.1$	&	$0.801 \pm 0.004$	&	216	\\	
CIV$\lambda4786$	&	$158.4 \pm 2.0$	&	$131.3 \pm 3.0$	&	$65.1 \pm 0.2$	&	$7190.56 \pm 0.39$	&	$0.206 \pm 0.020$	&	$78.53$	&	$17.3 \pm 1.2$	&	$0.927 \pm 0.021$	&	238	\\	
CIV$\lambda5018$	&	$383.7 \pm 1.7$	&	$154.3 \pm 2.6$	&	$104.6 \pm 0.9$	&	$7188.34 \pm 0.21$	&	$0.118 \pm 0.015$	&	$78.53$	&	$29.3 \pm 1.5$	&	$1.105 \pm 0.019$	&	238	\\	
CIV$\lambda5468$	&	$160.1 \pm 0.6$	&	$135.1 \pm 0.9$	&	$65.5 \pm 0.4$	&	$7188.86 \pm 0.09$	&	$0.326 \pm 0.005$	&	$78.53$	&	$17.0 \pm 0.3$	&	$0.922 \pm 0.006$	&	216	\\	
CIV$\lambda5804$	&	$529.0 \pm 0.9$	&	$119.5 \pm 1.5$	&	$68.8 \pm 6.2$	&	$7189.35 \pm 1.45$	&	$0.331 \pm 0.011$	&	$78.53$	&	$11.7 \pm 0.5$	&	$0.814\pm 0.011$	&	238	\\	
HeI$\lambda5876$	&	$205.2 \pm 0.5$	&	$158.8 \pm 0.9$	&	$107.4 \pm 0.3$	&	$7189.86 \pm 0.08$	&	$0.423 \pm 0.005$	&	$78.53$	&	$24.3 \pm 0.4$	&	$1.038 \pm 0.007$	&	238	\\	
HeII$\lambda4859$	&	$160.5 \pm 2.0$	&	$152.3 \pm 3.5$	&	$85.4 \pm 6.5$	&	$7184.28 \pm 1.97$	&	$0.379 \pm 0.021$	&	$78.53$	&	$22.8 \pm 1.7$	&	$1.017 \pm 0.025$ &	 204	\\	
HeII$\lambda5411$	&	$131.4 \pm 0.4$	&	$131.4 \pm 0.7$	&	$76.3 \pm 0.3$	&	$7190.55 \pm 0.07$	&	$0.338 \pm 0.004$	&	$78.53$	&	$15.4 \pm 0.2$	&	$0.892 \pm 0.005$	&	216	\\	
HeII$\lambda6560$	&	$738.9 \pm 0.4$	&	$137.7 \pm 0.6$	&	$118.1 \pm 0.3$	&	$7189.39 \pm 0.06$	&	$0.336 \pm 0.004$	&	$78.53$	&	$17.8 \pm 0.2$	&	$0.937 \pm 0.004$	&	216	\\	\hline
%O star
\multicolumn{10}{c}{O star absorption lines} \\ \hline
%	&		&		&		&		&		&		&		&		&			\\
H I + He II $\lambda4861$	&	$-13.5 \pm 0.9$	&	$79.6 \pm 1.4$	&	$252.7 \pm 0.3$	&	$7188.37 \pm 0.77$	&	$0.287 \pm 0.015$	&	$78.53$	&	$3.61 \pm 0.19$	&	$0.548 \pm 0.007$	&	370	\\	
HeI$\lambda5876$	&	$-22.8 \pm 1.4$	&	$65.2 \pm 2.7$	&	$293.4 \pm 4.0$	&	$7189.77 \pm 0.67$	&	$0.498 \pm 0.031$	&	$78.53$	&	$1.48 \pm 0.20$	&	$0.408 \pm 0.020$	&	492	\\	
HeII$\lambda5411$	&	$-5.6 \pm 4.0$	&	$63.3 \pm 6.3$	&	$267.6 \pm 26.4$	&	$7190.77 \pm 5.53$	&	$0.232 \pm 0.086$	&	$78.53$	&	$1.91 \pm 0.58$	&	$0.448 \pm 0.047$	&	370	\\	
\hline
\end{tabular}}
\end{table*}
%\end{landscape}

%Table: orbital_elements_table.ods

%
% \begin{figure*}
%\includegraphics[width=3in, angle=0]{vel_wr_plot.eps}
%\includegraphics[width=3in, angle=0]{vel_o_plot.eps}
%
%\caption{\label{figdynam} Orbital Fits....}
%
%\end{figure*}

\section{Photometric Modulation from the Colliding Winds}

\begin{figure*}
\includegraphics[angle=0,width=7in]{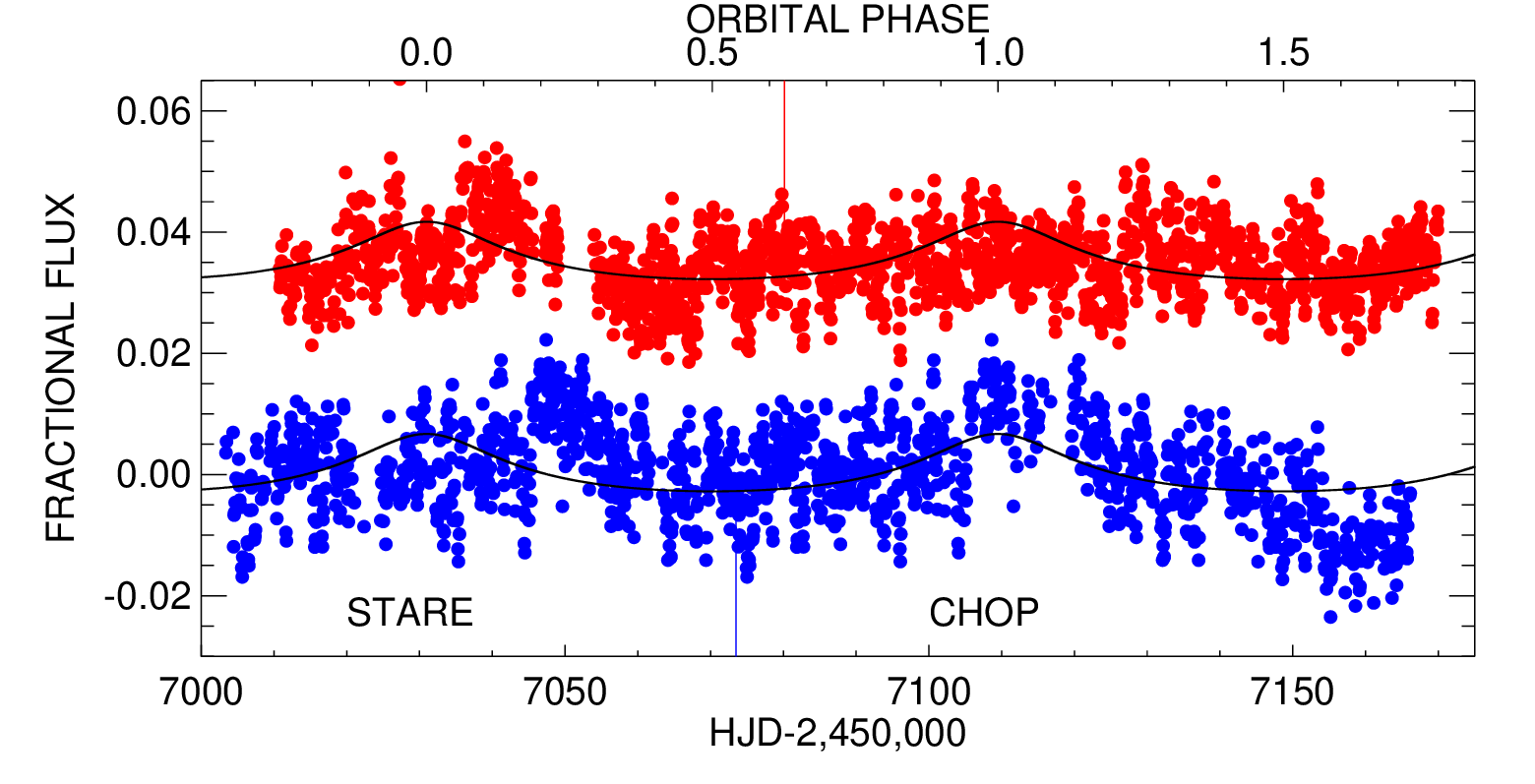}
\caption{The full light curve of \gv obtained with the \brite satellites, where each point shown represents the data binned onto the individual satellite orbits. The blue circles represent the blue data from {\it BRITE-Austria}, while the red points represent the photometry from {\it BRITE-Toronto}. The point where the normal mode was switched to a chopping mode is indicated by vertical dotted blue or red lines for the two satellites. The orbital phases are shown along the top x-axis, with time along the bottom x-axis. Errors on the individual points are smaller than the points, and the scatter will be discussed in a future analysis.}
\label{entireLC}
\end{figure*}

\begin{figure}
\includegraphics[angle=90,width=3.5in]{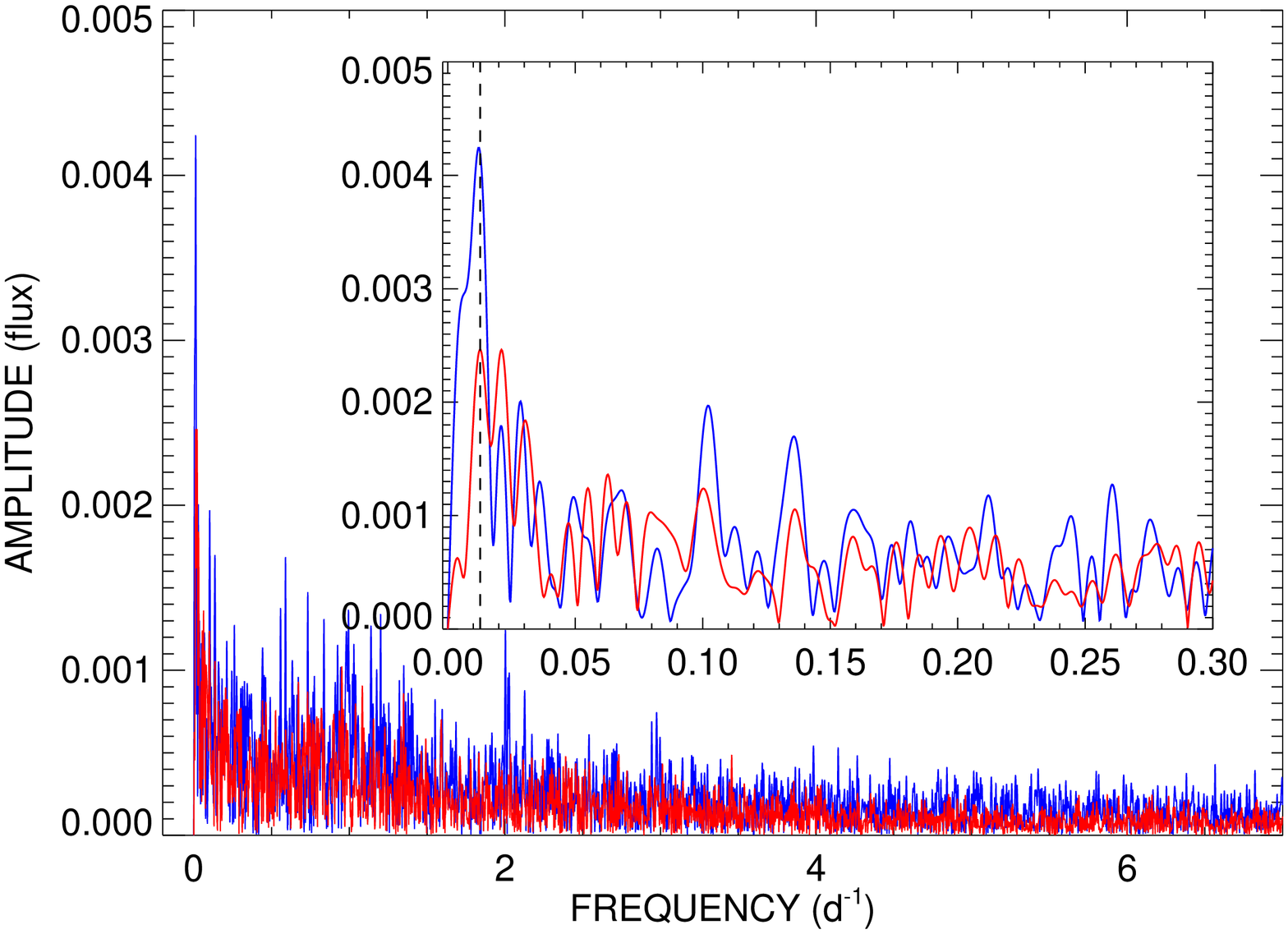}
\caption{Fourier transforms of the photometric data for {\it BRITE-Austria} (blue) and {\it BRITE-Toronto} (red). The inset panel shows the portion of the Fourier transform with longer periods highlighted. The vertical dashed line is the orbital frequency, which is the strongest frequency in both photometric datasets. }
\label{FT}
\end{figure}

In Figure \ref{entireLC}, we show the entire \brite light curve of \gv obtained in 2014--2015 with units of fractional flux. This measurement shows the change from the average value of the light curve, and a value of $+0.01$ indicates a 1\% increase in flux. The blue and red data correlate well with each other, and there is a general agreement between the data obtained with the two satellites. The most striking feature in these light curves is a long-term variation with a time-scale that is similar to the orbital period. We used {\tt Period04} \citep{2005CoAst.146...53L} to begin our analysis. The Fourier transforms of the photometric data are shown in Figure \ref{FT}. The Fourier transforms show similar peaks, confirming the visual similarities of the data from the two satellites. The peak is stronger in the blue filter, whereas the red peak seems to have a secondary peak at twice the orbital frequency. The frequencies are identical within the errors.

The time-series analysis shows that the strongest frequency present in the photometric data corresponds to the orbital frequency. The photometry from both {\it BRITE-Austria} and from {\it BRITE-Toronto} show a periastron peaking flux for the periastron passages, which is shown in Fig.~\ref{entireLC}. %The fit to these data seemed to be more complicated than a simple sinusoidal wave, which is likely expected in eccentric systems. Therefore, we did not fit the light curve with a simple sine wave to model the orbital variability.

\begin{figure}
\includegraphics[angle=90,width=3.0in]{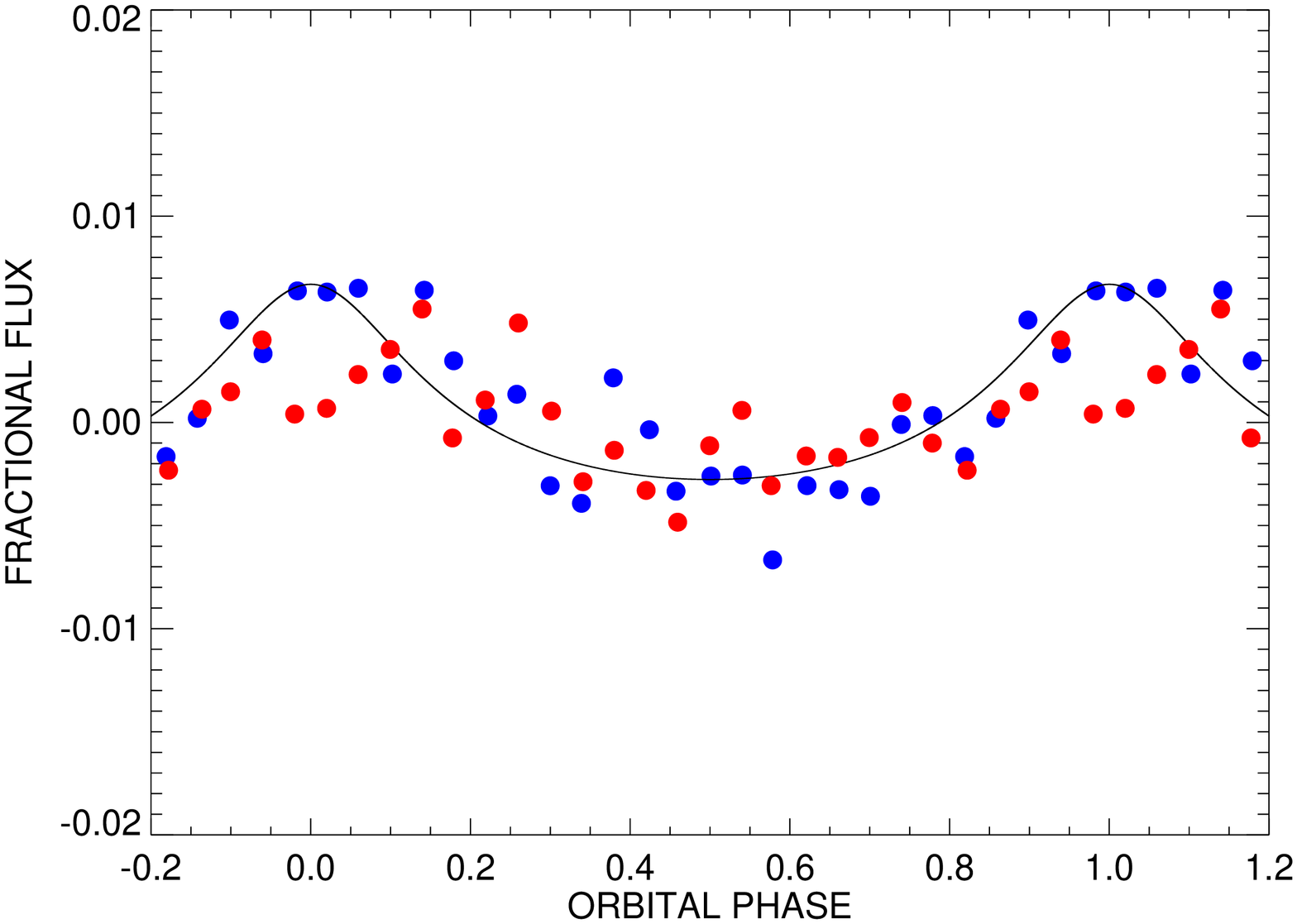}
\caption{The \brite light curves were phase-binned and examined for orbital variability. The resulting light curve shows an excess from the colliding winds with an inverse distance ($D^{-1}$) dependence. Error bars are smaller than the points.}
\label{BRITE-CW}
\end{figure}

In order to better characterize the data on the orbital time-scale, we binned the data in phase bins. Each bin had more than one hundred individual points used to derive the measured flux, making the relative errors negligible in the phased light curve presented in Figure \ref{BRITE-CW}. In this figure, there is a small amount of flux deficiency near periastron in the red compared with the blue. We fit the excess flux with a simple relation inversely proportional to the distance ($D^{-1}$) between the stars in their eccentric orbit. The $D^{-1}$ relationship is predicted for X-ray variations related to adiabatic cooling in CW systems, so one may expect similar results in the optical continuum \citep[e.g.,][]{2011MNRAS.418....2F}. The result is a reasonable fit of the excess flux, with an amplitude of 0.009 in the fractional flux. We have also overplotted this trend in the entire \brite light curve shown in Fig.~\ref{entireLC}

We could not find any other examples in the literature for a $D^{-1}$ photometric relationship in WR+O colliding winds binaries, but this is likely similar to the spectral line variability seen in other colliding winds systems, such as WR 140 \citep{2011MNRAS.418....2F}. In the case of WR 140, the eccentricity is much higher ($e=0.894$), and the spectral line variability follows a $D^{-2}$ dependency rather than a $D^{-1}$. Such a relation likely shows that the gas is fairly radiative in its cooling rather than adiabatic \citep{2011MNRAS.418....2F}. For our observations of $\gamma^2$ Vel, the difference between the $D^{-2}$ and $D^{-1}$ fits is minimal as the eccentricity is much lower ($e=0.33$), so we have adopted the adiabatic ($D^{-1}$) relation here. A similar photometric behavior is possibly seen in the highly eccentric system R 145 in the LMC, but the data obtained were sparse and not modeled in that instance \citep{2016arXiv161007614S}. Lastly, there was photometric modulation on the 208 d orbit of WR 25 seen in the ASAS photometry reported by \citet{2014ApJ...788...84P} with the photometric peak occurring near phase 0, but the light curve appears to be non-symmetric about the periastron passage.
.

\section{Spectroscopy of the Colliding Winds}

\begin{figure*}
\includegraphics[angle=90,width=6.in]{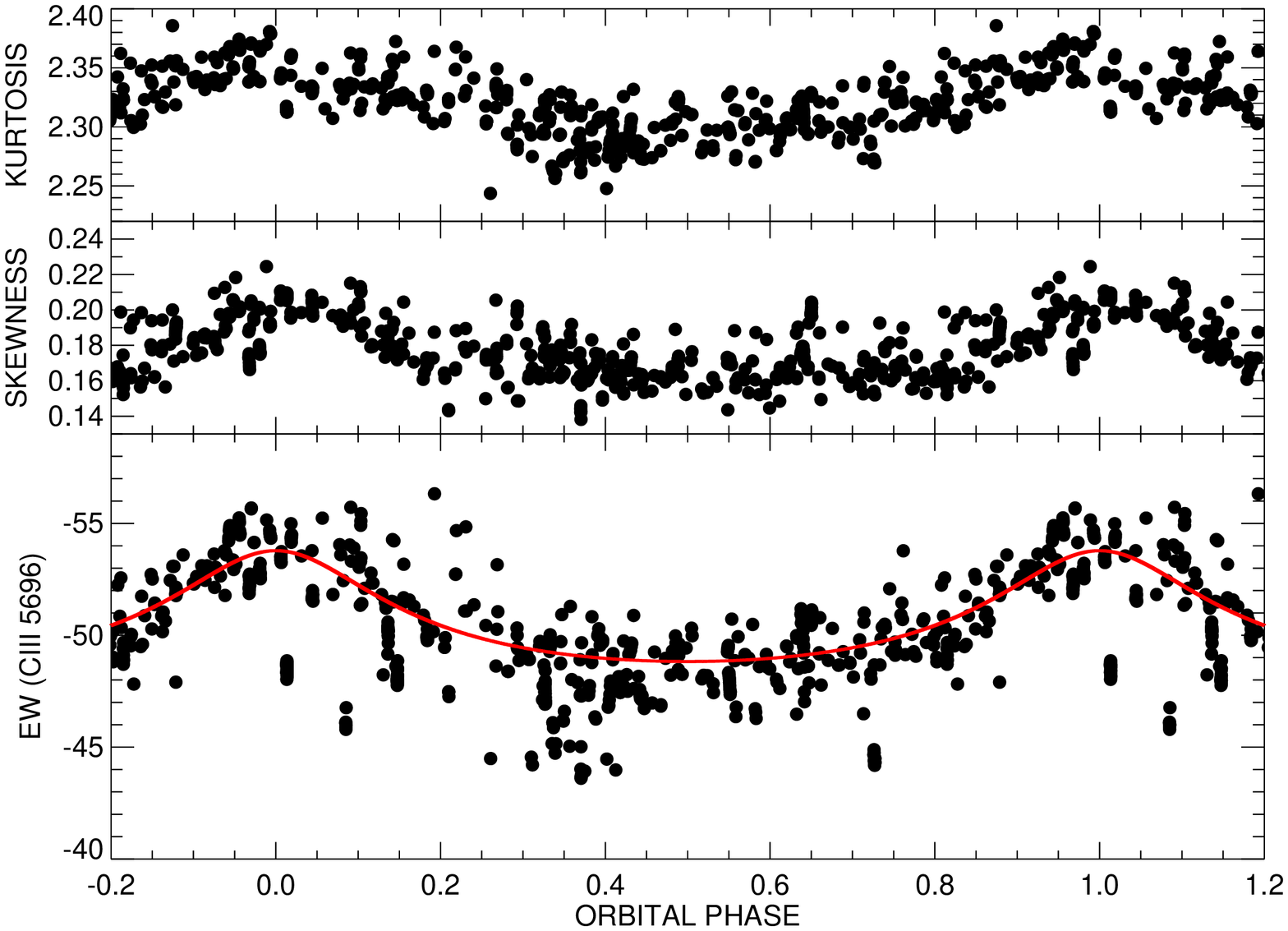}
\caption{Colliding winds excess in CIII measurements. Error bars are typically smaller than the points. The red line fit for the equivalent width represents a $D^{-1}$ fit, with an amplitude of 4.95$\pm0.10$ \AA.}
\label{CIII-CWmeasure}
\end{figure*}

Most of our spectroscopy focused on the C~III $\lambda 5696$ transition due to the atomic processes mentioned in Section 2.2. We began our analysis of the colliding winds of the system by measuring the equivalent width along with the normalized third and fourth central moments (skewness and kurtosis, respectively) of the line. (In a Gaussian distribution, the skewness is zero, and the kurtosis is three.) These measurements are shown in phase in Fig.~\ref{CIII-CWmeasure}, and show that the profile is always slightly skewed to the positive and that the profile has a kurtosis less than three, indicating that the light profile has a velocity distribution that reaches zero with a steeper slope than a Gaussian.

The behavior of the equivalent width of the C III line can be explained in terms of the colliding winds. Close to the periastron passage, the wind-wind collision is at a peak which falls off as the distance increases. As in the case of the photometric behavior (see previous section), we fit the variations with a $D^{-1}$ dependency, which is overplotted in Fig.~\ref{CIII-CWmeasure}. The measurements of the skewness and kurtosis peak near the periastron passage. The profile becomes the most non-Gaussian in terms of skewness near periastron, while the kurtosis shows that the profile becomes more centrally-peaked.

%$$ FW_{\rm ex} = C_1 + 2 v_{\rm strm} \sin \theta \sqrt{1-\sin^2 i \cos^2(\phi - \delta \phi)} $$
%
%$$ RV_{\rm ex} = C_2 + v_{\rm strm} \cos \theta \sin i \cos(\phi - \delta \phi) $$
%
%To account for a elliptical orbit, we must replace $\phi$ by $\phi - [90^\circ - \omega]$, so that $\phi$ now represents the true anomaly with $\phi = 0$ corresponding to the periastron passage.

\begin{figure*}
\includegraphics[angle=0,width=3.0in]{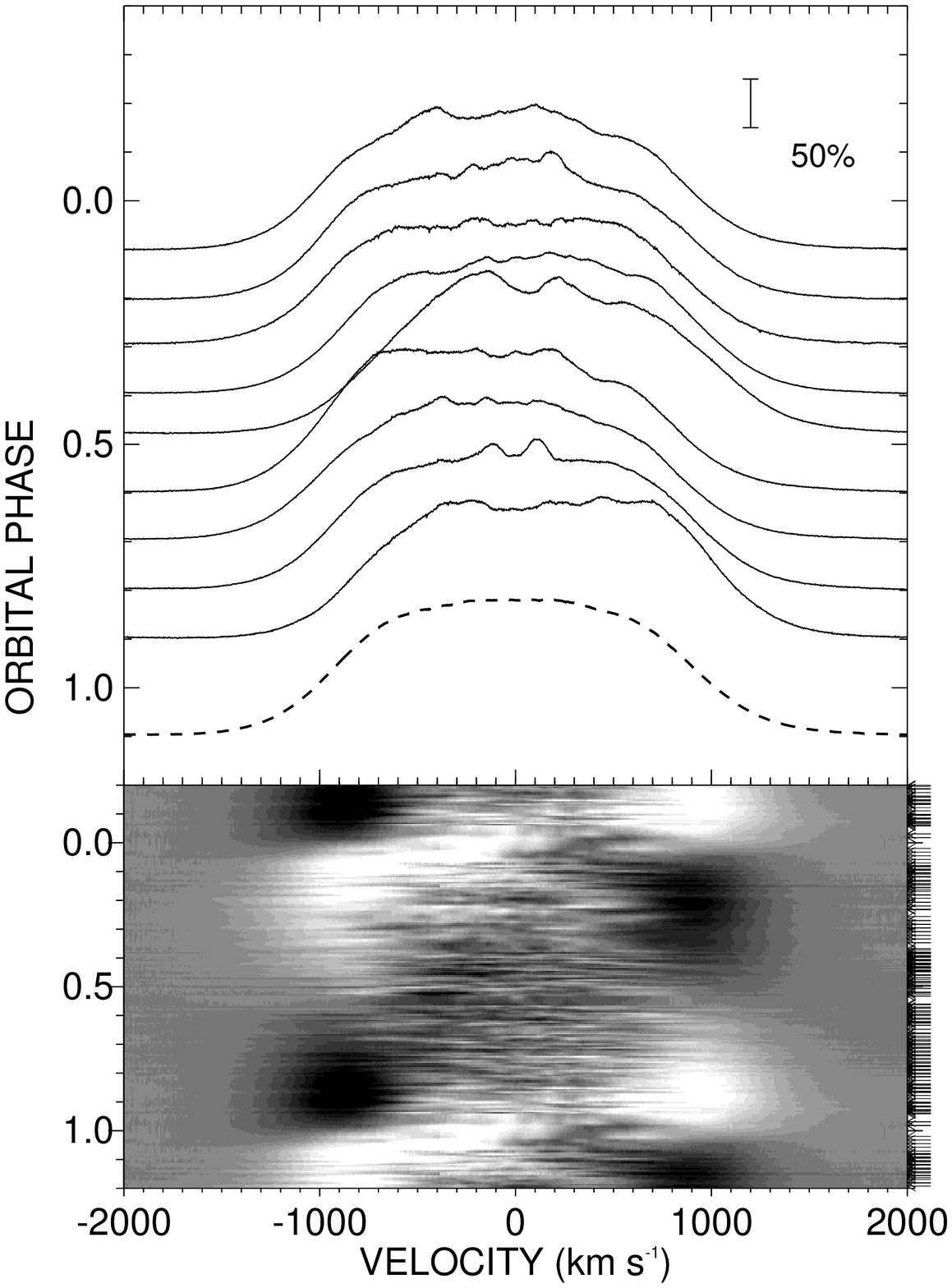}
\includegraphics[angle=0,width=3.0in]{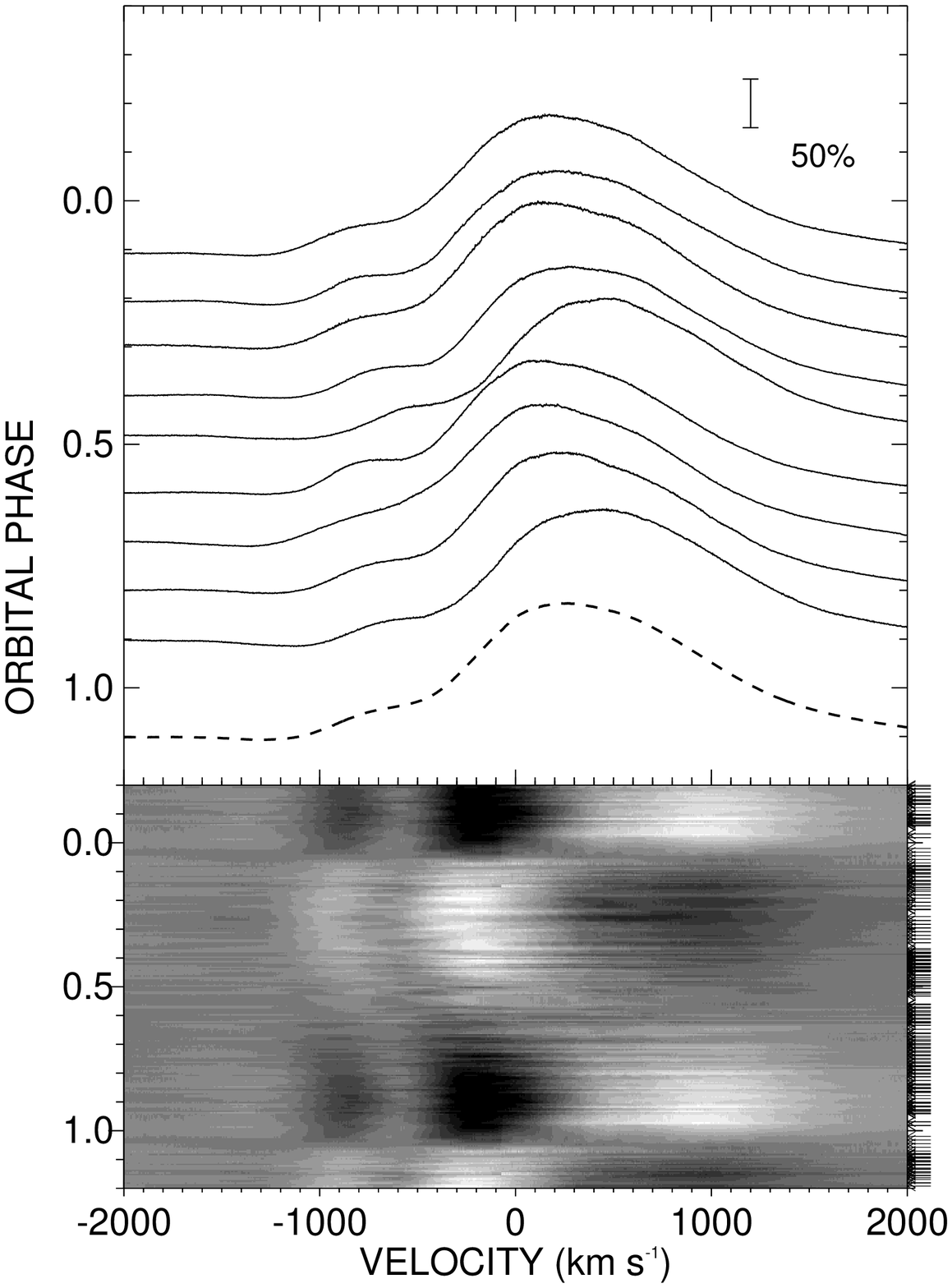}
\includegraphics[angle=0,width=3.0in]{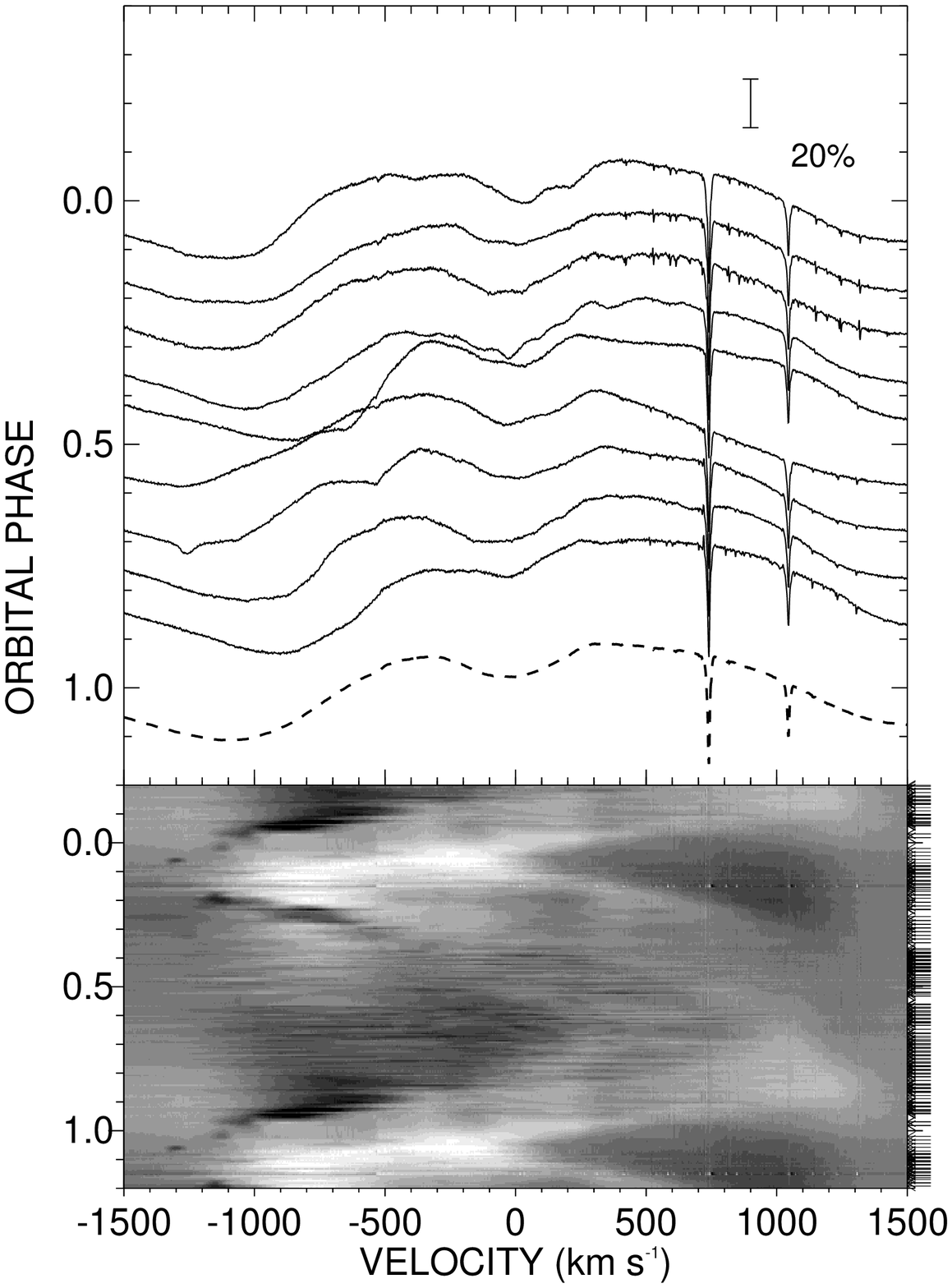}
\includegraphics[angle=0,width=3.0in]{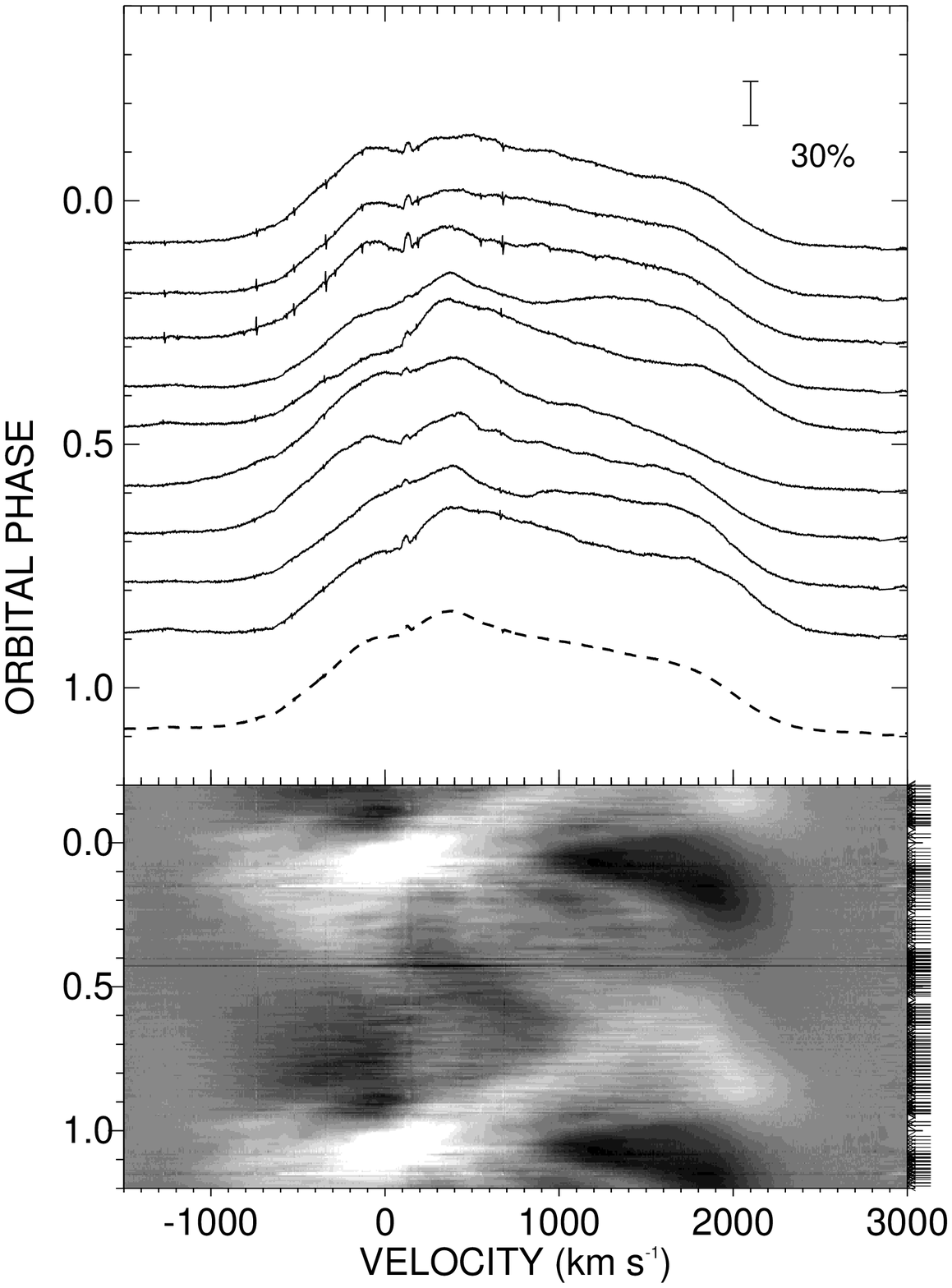}
\caption{CW excess for C III $\lambda 5696$ (top left), C IV $\lambda\lambda 5801, 5812$ doublet (top right), He I $\lambda 5876$ (bottom left), and He II $\lambda 6560$ (bottom right). In each case, we have subtracted the mean profile and only show the CHIRON spectra to maintain constant resolving power and similar S/N in the individual spectra. We also show the mean profiles that were subtracted in each plot, plotted with a thick dashed line.}
\label{specCW}
\end{figure*}

In Fig. \ref{specCW}, we show phased dynamical spectra of four spectral lines of varying sensitivity to the colliding-winds effects. Each sub-panel shows both the line spectra and a grey-scale representation of the spectra with the average profile subtracted from the observations. Each line shows a peak in emission at phase zero (periastron), as expected from the results of the equivalent width measurements in Fig.~\ref{CIII-CWmeasure}. The measurements of C III $\lambda 5696$ show that the skewness of the profile is fairly stable between phases 0.2 and 0.8, which is visually seen in these dynamical plots in the form of small amounts of variability during these phases.

In the He I $\lambda 5876$ profile, we note a striking appearance of a narrow P~Cygni type absorption.
It appears around phase 0.85 at $\sim -500$ km s$^{-1}$, accelerating to $-$1100 km s$^{-1}$ at phase 0.95, when it disappears.  At phase 0.20 it re-appears at $-$1100 km s$^{-1}$ and decelerates back to $-$500 km s$^{-1}$ before disappearing again until phase 0.85. We suspect that this absorption is related to the structure of the cooler parts of the shock cone located further from the two stars, and will return to this in the following section.

%
% \begin{figure*}
%\includegraphics[width=3in, angle=0]{ph_ciii_ALL.eps}
%\includegraphics[width=3in, angle=0]{ph_civ_ALL.eps}
%
%\includegraphics[width=3in, angle=0]{ph_hei5876_ALL.eps}
%\includegraphics[width=3in, angle=0]{ph_heii6560.eps}
%
%\caption{\label{figdynam} Difference Dynamical Spectra are pretty. CIII, CIV, He I 5876 and He II 6560}
%\end{figure*}

\section{Discussion}

This paper has amassed a large and unique spectroscopic and photometric data set which we can use to better understand the variability of Wolf-Rayet stars and binary systems. We explore the physics of the orbit, the photometric modulation, and wind shocks in this section.

\begin{figure}
\includegraphics[angle=90,width=3.0in]{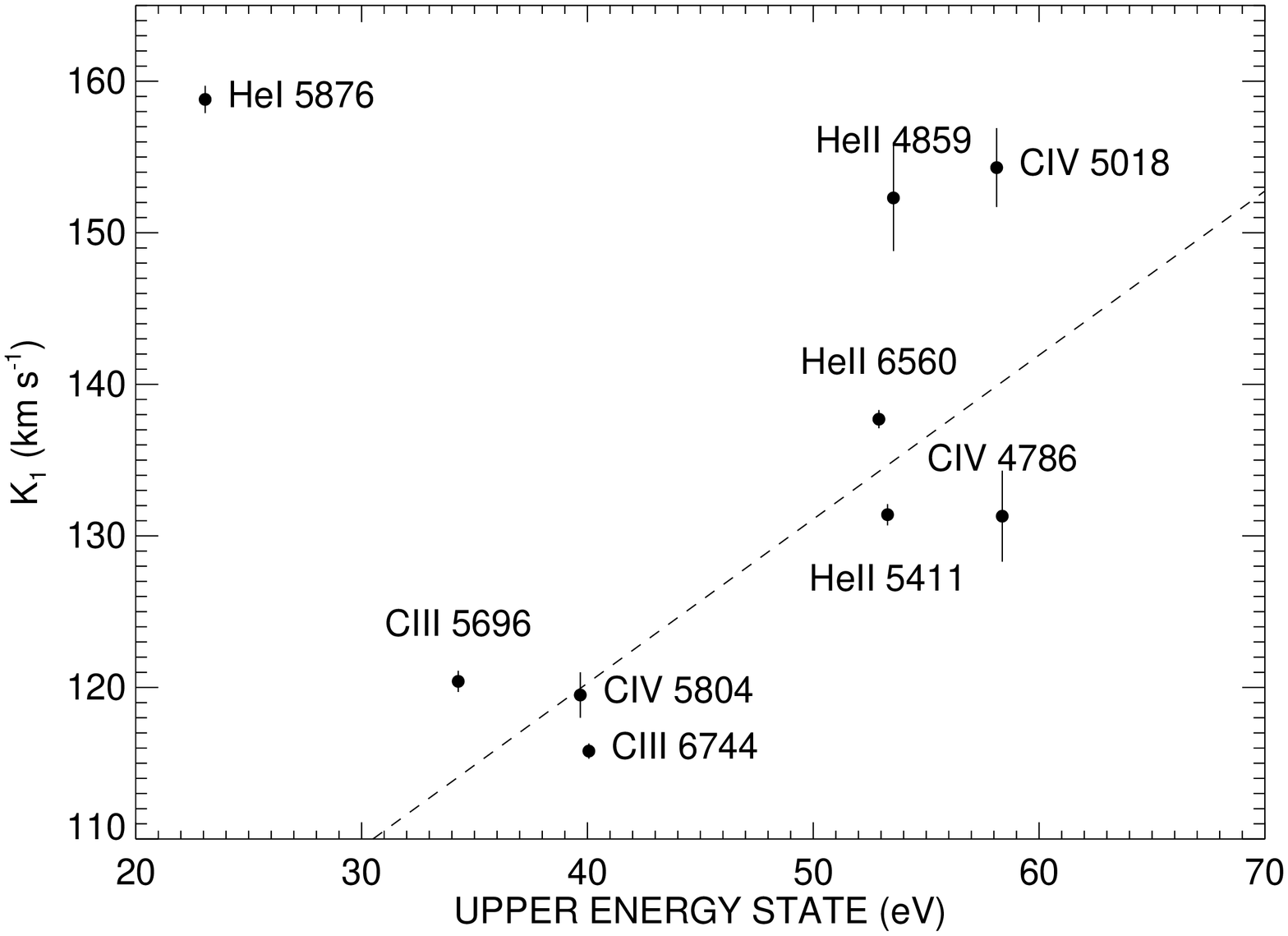}
\caption{Orbital semi-amplitude $K_1$ as a function of ionization potential.}
\label{Kdepend}
\end{figure}

\citet{2014AJ....148...62K} examined the massive SMC binary HD 5980, which consists of two early WNh stars. Despite the eclipsing nature of HD 5980, it has continued to be difficult to estimate the masses of the component stars. %\citet{2014AJ....148...62K} have done the best job to date to obtain reliable masses. 
In order to obtain masses, \citet{2014AJ....148...62K} elected to use the highest ionization lines that would form at the smallest radii from the WR star. In Fig.~\ref{Kdepend}, we show the effects of differing ionization potential on the observed semi-amplitude of the WC orbit in the \gv system from the orbital elements in Table \ref{table-elements} and the atomic data in the NIST database \citep{nist}.

The He~I $\lambda 5876$ orbit does not fit within the trend shown in Fig.~\ref{Kdepend}. He~I $\lambda$5876 is strongly influenced by absorption from the leading and trailing arms of the shock cone from the colliding winds in the system, which we think is the reason for the outlying point. The dependence between the upper energy state and the semi-amplitude is likely influenced by a few effects. First off, these lines often do show the effects of colliding winds, which manifest themselves as large emission perturbations that are sometimes double-peaked in nature. The perturbations could sway the overall velocity to the red or blue by a small amount, but mostly near the peak of the lines. These emission line velocities were measured near the base of the profile, so effects on these derived velocities from the central parts of the profile and peaks should be minimal.

\begin{figure*}
\includegraphics[angle=90,width=3.0in]{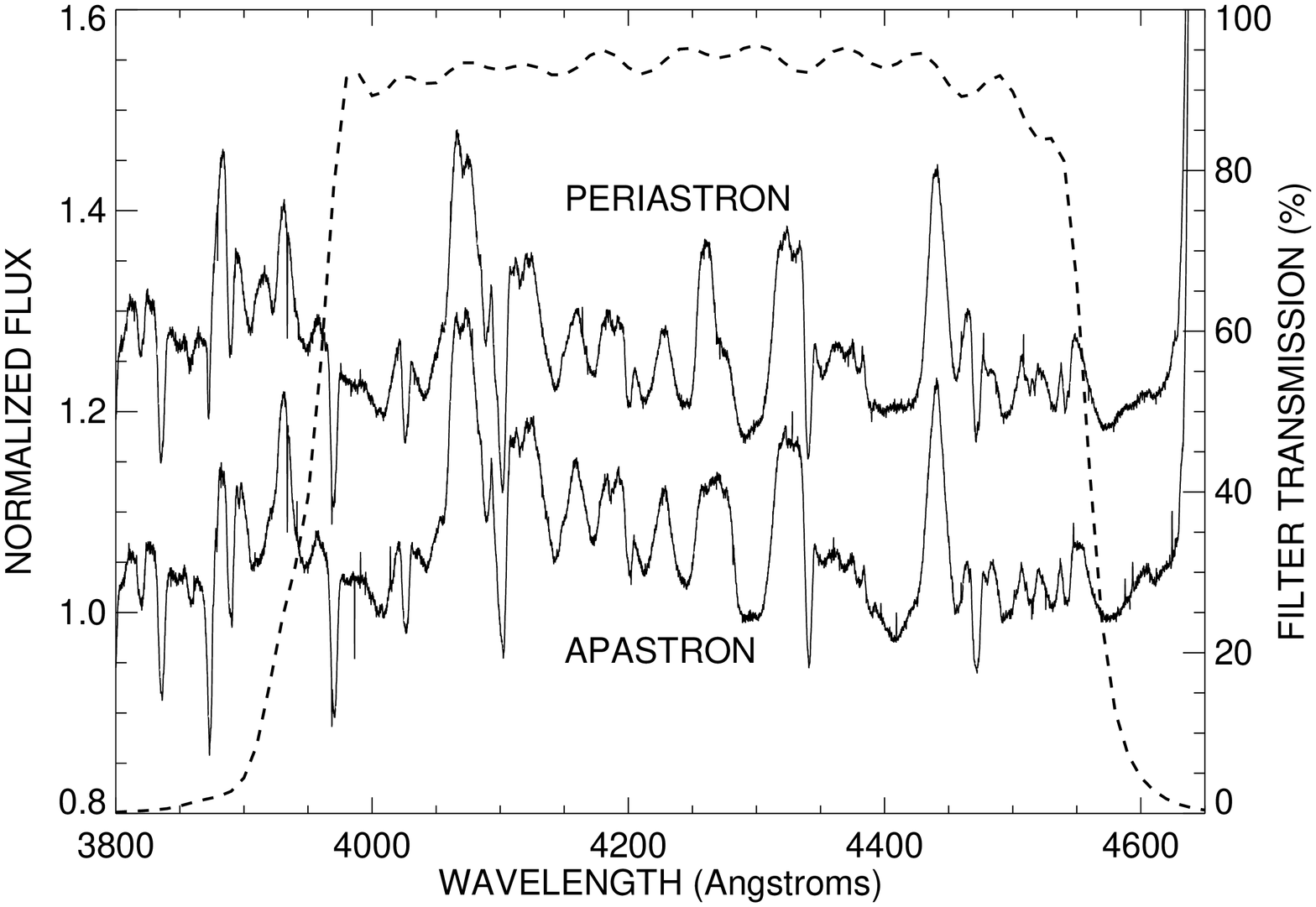}
\includegraphics[angle=90,width=3.0in]{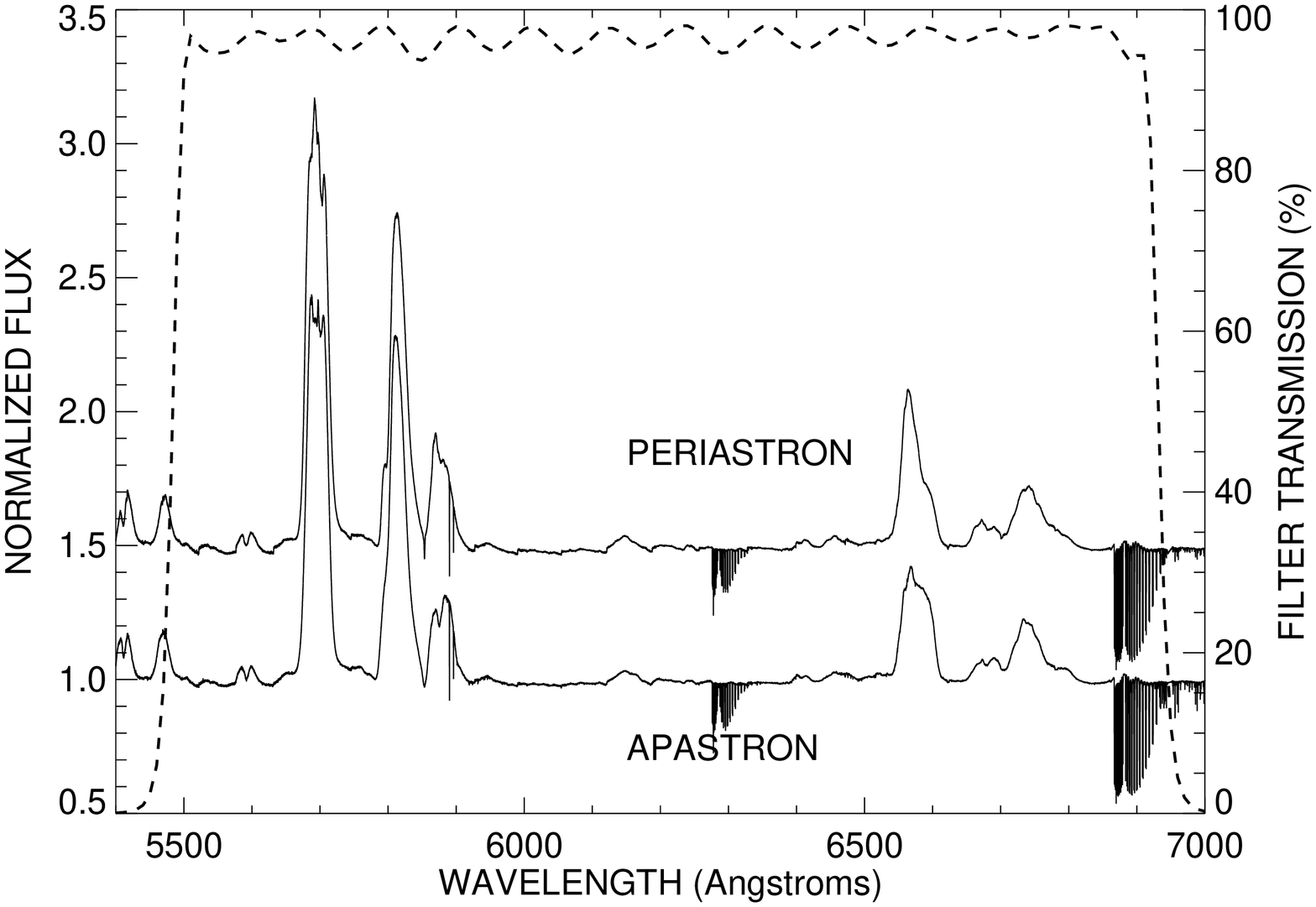}
\caption{BRITE filters compared to the spectrum of \gv at apastron and periastron. Spectra have been offset for clarity. While the differences are not obvious in the spectra, the integrated spectra across the {\it BRITE} filters shows the variations are comparable to that of the photometry, roughly 1\%.}
\label{filter-response}
\end{figure*}

The observational fact that a $D^{-1}$ trend is present for both the photometry (Fig. \ref{BRITE-CW}) and emission line spectroscopy such as the C III $\lambda 5696$ transition (Fig.~\ref{CIII-CWmeasure}) lends itself to questioning if the continuum is changing, or if the change is an effect of increased line production from the shocks at periastron. In order to determine the source of this effect, we compared blue spectra taken at apastron and periastron by \citet{1997A&A...328..219S} with the {\it BRITE} blue filter response. We also compared the {\it BRITE} red filter response with our new CHIRON spectroscopy (Fig. \ref{filter-response}). These results show that the variations, when integrated across the spectrum convolved with the filter response, are identical. In both cases, the continuum was normalized to unity, implying that the photometric changes are reflecting the spectroscopic changes from apastron to periastron, namely that a select few lines contribute a small amount of excess when the stars are closest in the orbit. Likely, these changes also explain the photometric excess seen in other eccentric WR binaries such as R145 \citep{2016arXiv161007614S}, which might be expected for a hot, optically-thin plasma.

We built smoothed particle hydrodynamic (SPH) simulations of the colliding winds in \gv, using the parameters given in Table \ref{table-SPHparams}.  The code originated from \citet{BenzP90} and \citet{BateBonnellPrice95}, with the first application to a colliding wind binary system by \citet{OkazakiP08}.  Current capabilities of the code are described by \citet{MaduraP13} and \citet{RussellP16}, which include radiative cooling of the shocked plasma and the acceleration of each wind from its stellar surface according to a $\beta-$velocity law, $v(r) = v_\infty(1-R/r)^\beta$, where $v_\infty$ is the terminal wind speed and $R$ is the stellar radius.
The wind speeds and mass-loss rates for the two stars were taken from \citet{1999A&A...345..163D} and \citet{deMarcoWR}. The orbital elements were a combination of the visual elements of \citet{North} and our radial velocity results presented here. In particular, we use our radial velocity orbits, but the inclination from \citet{North}. We note that these simulations can qualitatively be used, but require better X-ray data near periastron to constrain how to incorporate radiative braking \citep{GayleyOwockiCranmer97}. Nevertheless, these simulations provide an excellent framework within which to examine our spectroscopic results.

\begin{table}
\centering
\caption{ Stellar, wind, and orbital parameters of the SPH simulation of \gv. \label{table-SPHparams}}
\begin{tabular}{l c c c}
\hline \hline

Star & O7.5 & WC8 & Reference \\
\hline
%& O7 & WC8 & \\
$M$ ($M_\odot$) & 28.5 & 9 & \citet{North}\\
$R$ ($R_\odot$) & 17 & 6 & \citet{North} \\
$\log{\dot{M}}$ ($M_\odot$\,yr$^{-1}$) & $-$6.75 & $-$5 & \citet{North} \\
$v_\infty$ (km\,s$^{-1}$) & 2500 & 1550 & \citet{deMarcoWR}\\
$P$ (d) & \multicolumn{2}{c}{78.53}& \citet{North} \\
$a$ (AU) & \multicolumn{2}{c}{1.2}& this work \\
$e$ & \multicolumn{2}{c}{0.333} & this work\\

\hline \hline
\end{tabular}
\end{table}

\begin{figure*}
  \includegraphics[width=\textwidth]{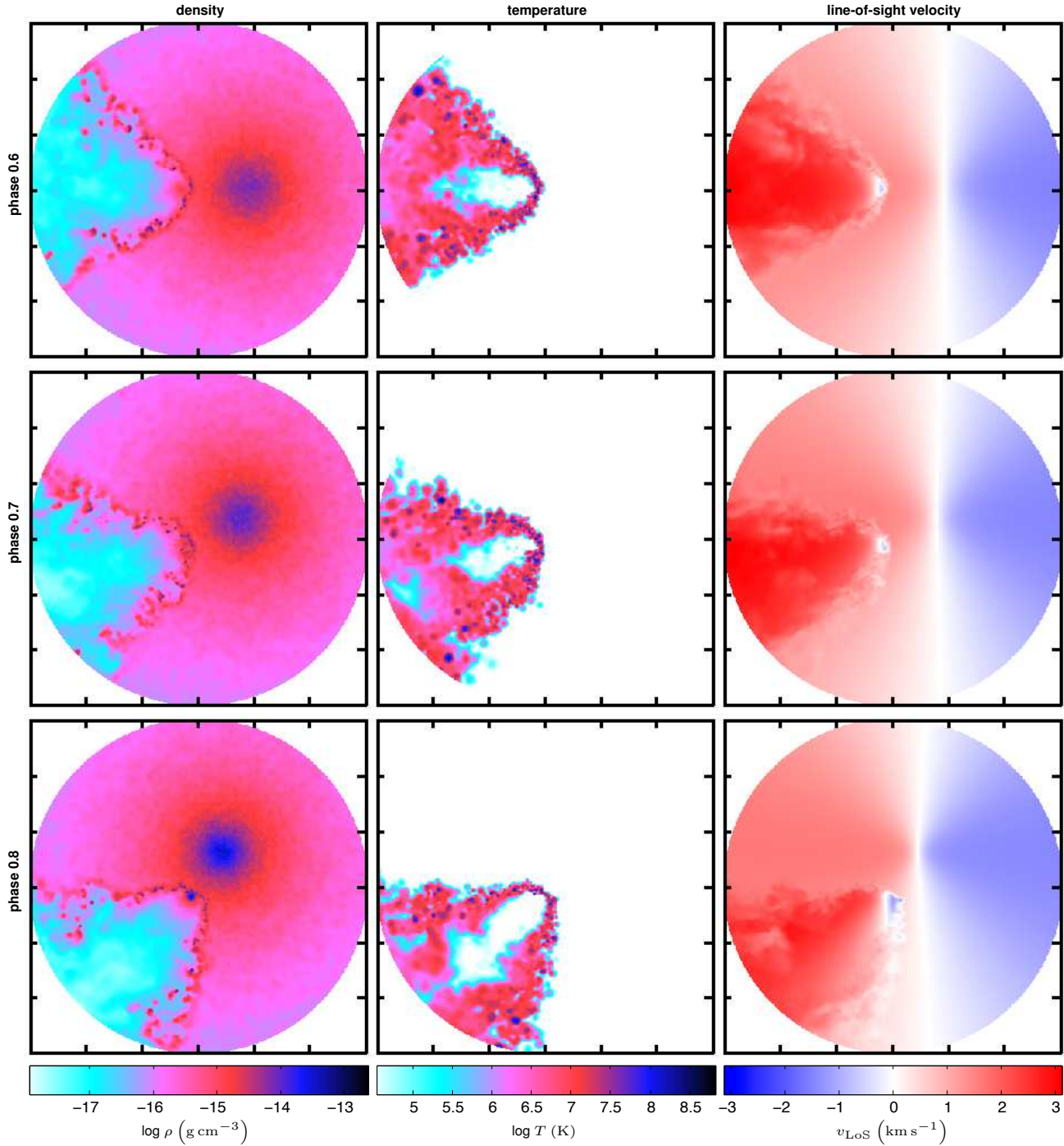}
  \caption{Density (left), temperature (center), and line-of-sight velocity (right) of the hydrodynamic simulation of the colliding winds in \gv.  The plane shown is rotated and inclined from the orbital plane such that the observer is directly to the right of the frame.  Each row shows a different phase; 0.6-0.8 for this figure. At phase 0.6, the WR star is on the right.  The orbital motion is counterclockwise.}
  \label{fi:SPH1}
\end{figure*}

%\clearpage

\begin{figure*}
  \includegraphics[width=\textwidth]{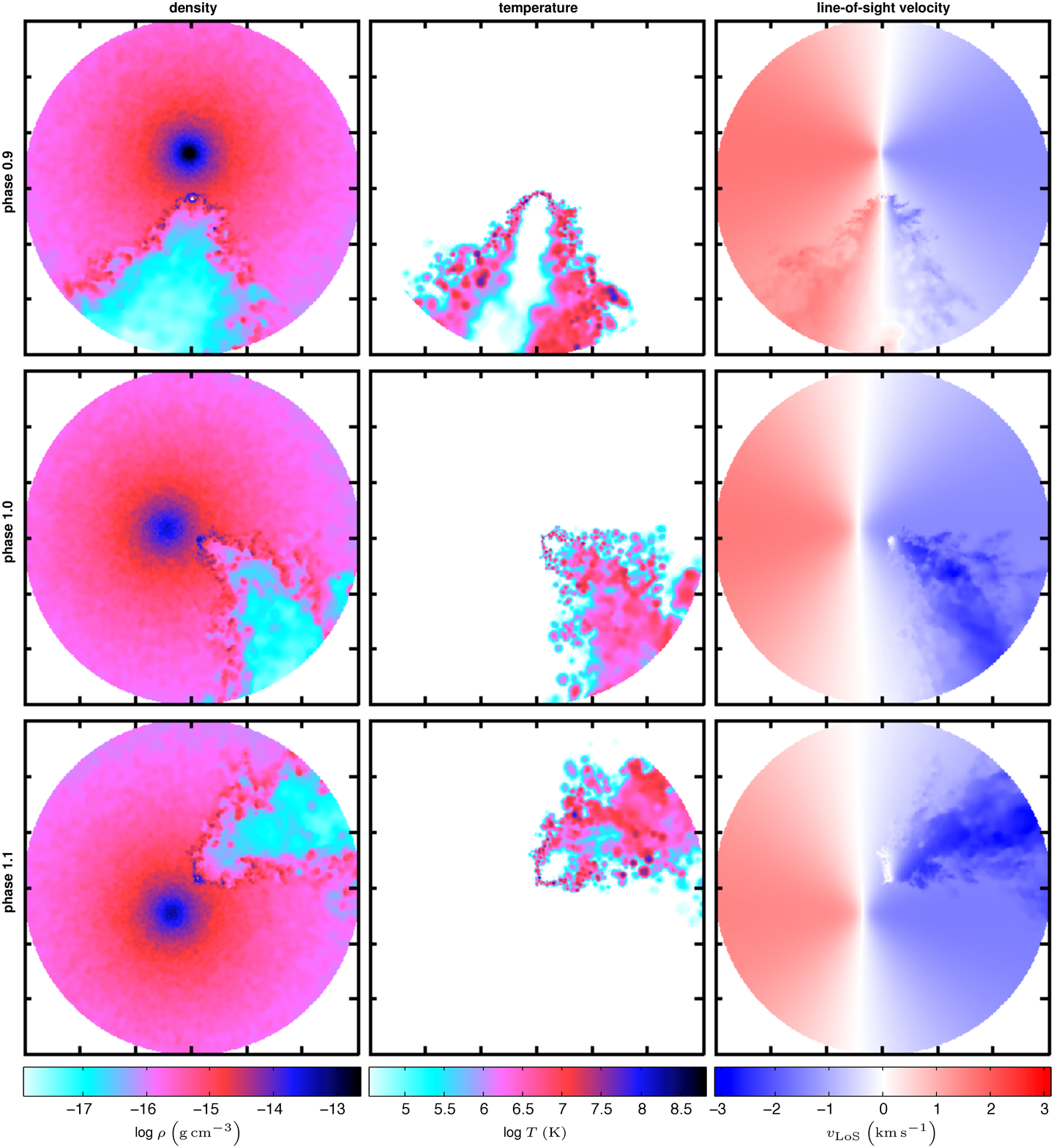}
  \caption{Same as Fig.~\ref{fi:SPH1}, but for phases 0.9-0.1. The observer is to the right, and the simulations are inclined to our observed geometry.}
  \label{fi:SPH2}
\end{figure*}

%\clearpage

\begin{figure*}
  \includegraphics[width=\textwidth]{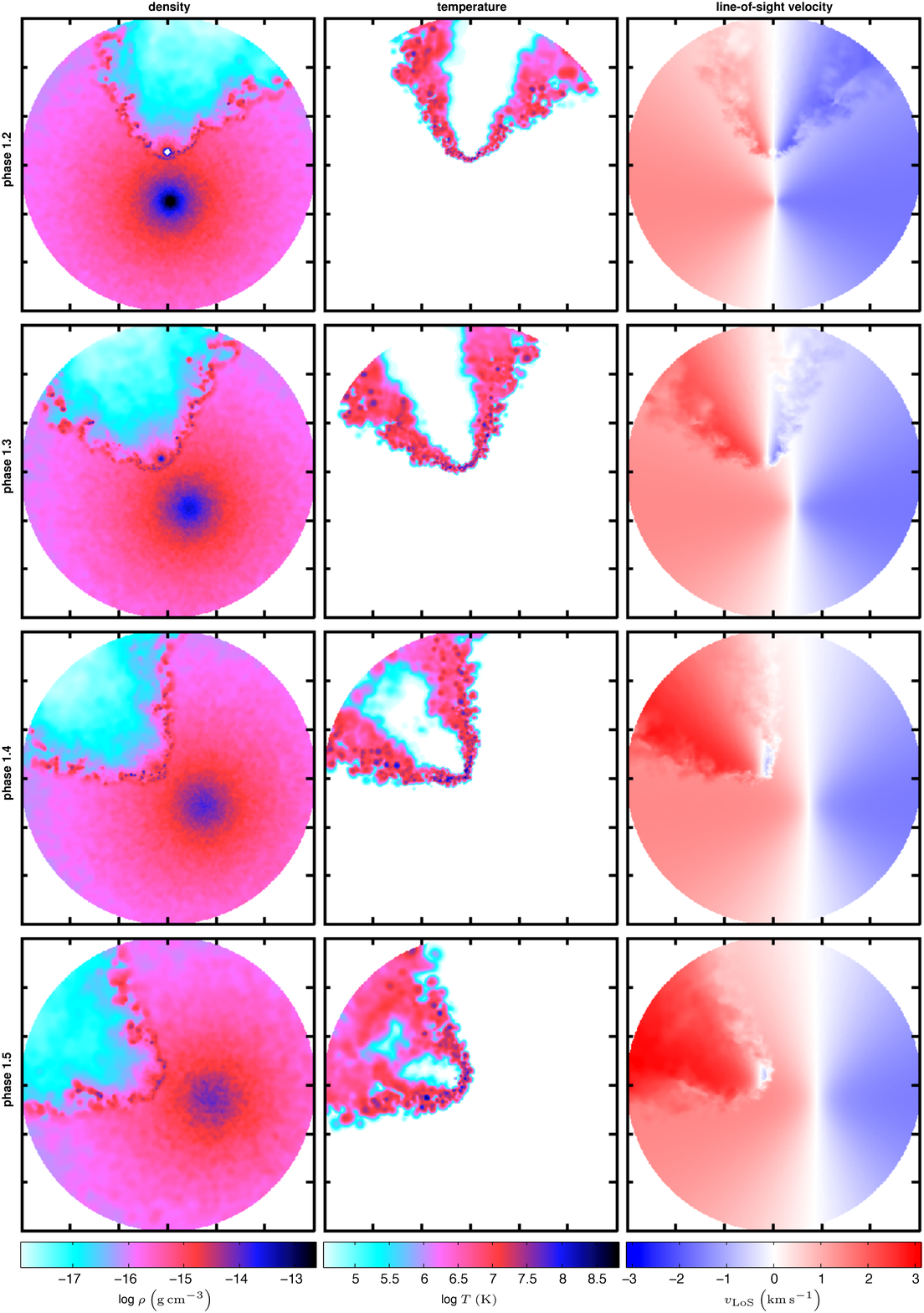}
  \caption{Same as Fig.~\ref{fi:SPH1}, but for phases 0.2-0.5. The observer is to the right, and the simulations are inclined to our observed geometry.}
  \label{fi:SPH3}
\end{figure*}

Figs. \ref{fi:SPH1}--\ref{fi:SPH3} show the density, temperature, and line-of-sight velocity of the SPH simulation through a full orbital cycle.  The particular plane shown is close to the orbital plane, but is rotated according to the orbital elements of \citet{North} and our own results to place the observer to the right of the frame. The WR star (in projection) is on the right of the top panel of Fig.~\ref{fi:SPH1}, and the orbital motion is counterclockwise.  We use these figures to describe features of the model in a qualitative sense.

%Figs. \ref{fi:SPH1}--\ref{fi:SPH3} show the density, temperature, and \sout{radial} velocity of the gas in our line of sight from the SPH simulation, where the observer is located along the positive $x$-axis. Using the orbital elements of \citet{North} and our own results, we place the observer along the positive $x$-axis. \sout{Each figure has been matched in the color scale, size, and orientation to allow for inter-comparison.} We use these figures to describe features observed with these models in a qualitative sense. %The figures are placed into our line-of-sight using the orbital elements from \citet{North} along with our results.

The first spectroscopic feature we will discuss is the excess emission from the C III $\lambda 5696$ line. In the dynamical spectra in Fig.~\ref{specCW}, we see that the line profiles exhibit an excess that moves across the profile during the orbit, with a strength in the excess that peaks at periastron (Fig.~\ref{CIII-CWmeasure}). This feature shows blue-shifted emission near phase 0.1, and red-shifted emission near phase 0.7. These features can be seen in the SPH figures. For example, in the top panel of Fig.~\ref{fi:SPH2}, we see a hotter part of the shock cone below the WR star. This feature also has a radial velocity centered near 0, and would be easily seen as the C III excess at that time. At phases 0.1 and 0.7 (bottom panel of Fig. \ref{fi:SPH2} and middle panel of Fig. \ref{fi:SPH1}), we see the most extreme blue and red shifts in the excess, which can again be attributed to formation a little ways downstream from the head of the shock cone.

\begin{figure*}
\includegraphics[angle=0,width=3.0in]{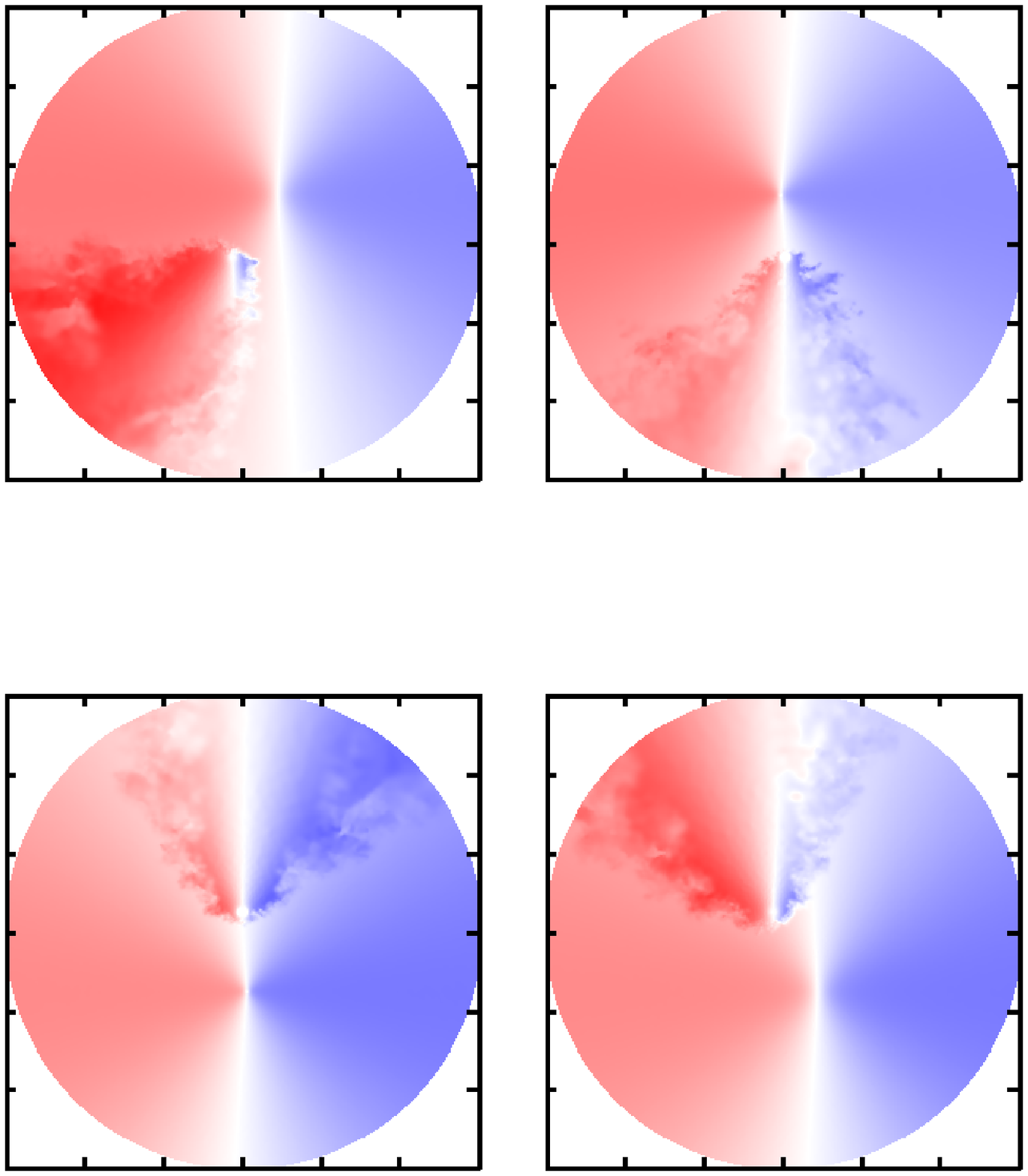}
\includegraphics[angle=0,width=3.0in]{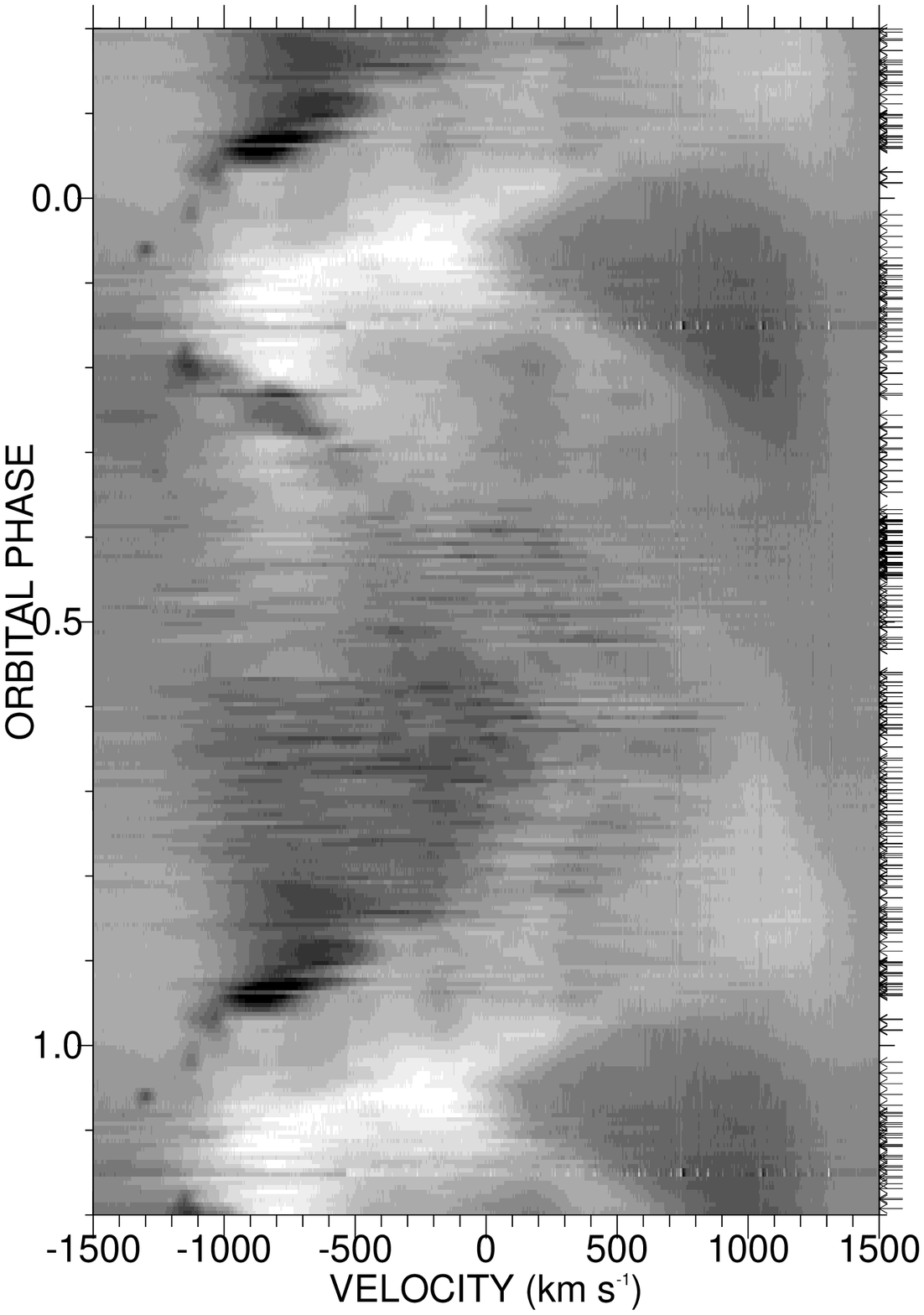}
\caption{The left panels depict the radial velocity from the SPH simulations at phases 0.8 (top left), 0.9 (top right), 0.2 (bottom left), and 0.3 (bottom right) in Figs.~\ref{fi:SPH1}--\ref{fi:SPH3}, while the right panel shows a larger view of the grey scale plot of He I $\lambda$5876 in Fig.~\ref{specCW}. We note that as the first shock arm crosses our line of sight in the He~I phased plot at phases $\sim$0.8--0.95, there is a strong blue-shifted density enhancement seen in the shock arm. Similarly at phases 0.2--0.3, the trailing arm crosses our line of sight with similar blue-shifted radial velocities. }
\label{HeI_SPH}
\end{figure*}

Another feature that is easily explained by the SPH simulations is the He I excess P Cygni absorption trough (Fig.~\ref{specCW}). This absorption feature shows up around $-500$ km s$^{-1}$ near phase 0.8, and rapidly accelerates to $-1000$ km s$^{-1}$ near periastron. A deviation in the blue portion of the radial velocity plots for the bottom panel of Fig. \ref{fi:SPH1} and the top panel of Fig. \ref{fi:SPH2} (phases 0.8 and 0.9, respectively) shows that some gas in the shock cone has a strong blueward radial velocity trend. These parts of the shock cone, in the leading arm, are likely cooled sufficiently to have neutral helium. Thus, they are likely the source of the excess P Cygni absorption. We show the relevant plots again in Fig.\ \ref{HeI_SPH} for clarity. 

We expect that the points in this plot where the lowest accelerated velocities of the cone appear and disappear provide a basis for the shock cone opening angle for the system (Fig.~\ref{specCW}, bottom left panel). As these points are separated by about 0.4 of the orbital phase (phases $\sim 0.85$ and $0.25$), this would translate to an angle of roughly 70$^\circ$. This can be broadly supported with a L\"{u}hrs model, as described by \citet{2000MNRAS.318..402H}. In the approximation, the radial velocity amplitude is proportional to both $\cos \theta$ and $\sin i$, where $\theta$ is the half-opening angle of the cone and $i$ is the inclination. We know the inclination thanks to the interferometry of \citet{North} and \citet{2017arXiv170101124L}, where $i=65.5^\circ$. As the semi-amplitude is $\sim$half of the width of the emission line, we should expect the opening angle to be near 60--65$^\circ$. We plan to make a more detailed model in a future analysis of these spectra.

The He I absorption returns with a high velocity near phase ~0.2 at the similar high velocity of $-1000$ km s$^{-1}$, and then propagates to lower velocities ($-500$ km s$^{-1}$) by phase 0.3 (see Fig.~\ref{HeI_SPH}). This second absorption feature is less strong, and would be formed in the trailing arm of the shock cone. A feature similar to the leading arm in the bottom panel of Fig. \ref{fi:SPH1} at phase 0.8 is readily seen at the SPH figure corresponding to phase 0.3 (second panel of Fig. \ref{fi:SPH3}). While a small deviation appears at phase 0.4 in the SPH figures, we expect that this is either absent or barely noticeable in the He I spectroscopy due to the close proximity of the gas to the binary, where the gas has a higher temperature. A small excess of C III emission is observed at the same phase and velocity, further strengthening this argument. Similar features are seen in other CW binaries, such as V444 Cygni \citep{1994ApJ...422..810M, 1997ApJ...485..826M}.

The SPH simulation is extremely helpful in qualitatively describing the variability present in the system. We will improve upon the SPH simulations in the near future with the addition of new X-ray spectra from {\it XMM-Newton}. Furthermore, we will examine the quantitative behavior of the C III excess with the modeling tools used by \citet{2000MNRAS.318..402H} to further constrain the system.

\section{Conclusions and Future Work}

This study presents the first look into a large dataset incorporating a six-month high-precision light curve from \brite and about 500 high-resolution optical spectroscopy. We confirmed the double-lined spectroscopic orbital elements of \citet{North}, which also confirms that the current masses are $M_{\rm O} = 28.5 M_\odot$ and $M_{\rm WC} = 9.0 M_\odot$. The colliding winds produce measurable photometric and spectroscopic effects. Lines such as C~III $\lambda$5696 show an equivalent width variation inversely proportional to the separation of the two stars in the system. Similar spectral variability is likely present across the optical spectrum, as the photometric variability can be reproduced with a $D^{-1}$ trend. Integration of the spectra convolved with the \brite filters shows that the trend can be reproduced with only excess line emission.

Our SPH models are consistent with the models presented by \citet{2017arXiv170101124L}, and are able to qualitatively describe the spectroscopic effects of the colliding winds in the optical emission lines. Similarly, the effects of the emission and absorption variability in the $K$-band were described in context of the numerical simulations and interferometric measurements of \citet{2017arXiv170101124L}. \gv was observed with {\it BRITE} again in late 2016 and early 2017. In parallel with these observations, we have collected X-ray spectroscopy with {\it XMM-Newton} across a periastron passage. The X-ray observations will provide the necessary constraints for radiative braking to improve our SPH modeling efforts and provide diagnostic information about the wind collisions. 

These results are the beginning of an in-depth look into the closest, brightest colliding-winds binary, $\gamma^2$ Velorum. We plan on performing a detailed analysis of the short-term variations in the light curve and spectroscopy in the near future. Further, we intend to model the optical spectroscopy of C III $\lambda 5696$ and other lines in the context of colliding winds in the same manner as \citet{2000MNRAS.318..402H}.
%Many light curves of WR stars were obtained with the {\it MOST} satellite. In general, these light curves show coherent, non-periodic signals over time scales of days. Similar variations are seen in the  {\it BRITE} light curve of $\gamma^2$ Velorum, but unlike the {\it MOST} WR targets, we amassed a very large collection of high-quality spectroscopy concurrent with the photometry. As such, we are now in a position to best interpret these data and then interpret the physics of the variability of WR stars in general.

%We have begun working on modeling  Unlike every other colliding-wind binary except for WR 140, we have a very high quality spectroscopic and visual orbit. As such, the variations observed here will yield strong constraints into the modeling efforts for colliding winds, as we know a priori the inclination angle of the system from the visual orbit, as well as from polarimetry \citep[$i\sim 70^\circ$;][]{1987ApJ...322..870S}.

%Our Further, these observations will probe if optical and X-ray variability is correlated for CW binaries such as $\gamma^2$ Velorum.

\section*{Acknowledgements}
This work is based in part on data collected by the {\it BRITE-Constellation} satellite
mission, built, launched and operated thanks to support from the
Austrian Aeronautics and Space Agency and the University of
Vienna, the Canadian Space Agency (CSA), and the Foundation for
Polish Science \& Technology (FNiTP MNiSW) and National Science
Centre (NCN).
We thank Werner Schmutz and Orsolo de Marco for providing the spectra that were used in their previous studies of the system. The Polish participation in the BRITE project is secured by NCN grant 2011/01/M/ST9/05914 and a SPUB grant from the Polish Ministry of Science and Higher Eduction. This work was based in part on observations at Cerro Tololo Inter-American Observatory, National Optical Astronomy Observatory (NOAO Prop. ID: 2015A-0133; PI: N. D. Richardson), which is operated by the Association of Universities for Research in Astronomy (AURA) and the SMARTS Consortium under a cooperative agreement with the National Science Foundation. This work also uses observations made from the South African Astronomical Observatory (SAAO). The results are based in part on spectroscopic data obtained through the collaborative Southern Astro Spectroscopy Email Ring (SASER) group. The professional authors of this paper are grateful to the amateur astronomers of the SASER team, who invested personal time and a contagious enthusiasm for this project. Further resources supporting this work were provided by the NASA High-End Computing (HEC) Program through the NASA Advanced Supercomputing (NAS) Division at Ames Research Center.

NDR acknowledges postdoctoral support by the University of Toledo and by the Helen Luedtke Brooks Endowed Professorship, and is thankful for his former CRAQ (Qu\'ebec) fellowship which supported him during the early phases of this project. LSJ was supported through an NSERC bursary for undergraduate research. 
CMPR acknowledges support from Chandra Theory Grant TM7-18003Z, which in turn supports his appointment to the NASA Postdoctoral Program at the Goddard Space Flight Center, administered by Universities Space Research Association, through a contract with NASA.
%CMPR is supported by an appointment to the NASA Postdoctoral Program at the Goddard Space Flight Center, administered by the Universities Space Research Association through a contract with NASA. 
AFJM and NSL are grateful for financial aid from NSERC (Canada) and FQRNT (Quebec). TS is grateful for financial support from the Leibniz Graduate School for Quantitative Spectroscopy in Astrophysics, a joint project of the Leibniz Institute for Astrophysics Potsdam (AIP) and the institute of Physics and Astronomy of the University of Potsdam. A. Popowicz acknowledges Polish National Science Centre grant: 2016/21/D/ ST9/00656. RK and WW acknowledges financial support by the Austrian Research Promotion Agency (FFG). 
GH acknowledges support from the NCN grant 2015/18/A/ST9/00578. AP acknowledge support from the Polish National Science Centre grant No. 2016/21/B/ST9/01126. TR acknowledges support from
the Canadian Space Agency grant FAST. GAW acknowledges Discovery Grant support from the Natural Sciences and Engineering Research Council (NSERC) of Canada.

%We thank Douglas Gies for discussions related to velocity measurements and spectral disentangling that were very helpful.

\bibliographystyle{mnras}
\bibliography{paper1_orbit_v2}

%\begin{thebibliography}{99}
%\bibitem[Lenz \& Breger(2005)]{period04}Lenz P., \& Breger M. 2005, Commun. Asteroseismol., 146, 53
%\bibitem[NIST(2016)] Kramida, A., Ralchenko, Yu., Reader, J. and NIST ASD Team (20l6). NIST Atomic Spectra Database (version 5.4), Available: http://physics.nist.gov/asd
%\end{thebibliography}

%
%
%\appendix
%\section{Supplementary figures}
%\clearpage
%
%\begin{figure*}
%\includegraphics[angle=0,width=7.0in]{vel_wr_plot.eps}
%\caption{WR star orbit }
%\label{WRorbit}
%\end{figure*}
%
%\begin{figure*}
%\includegraphics[angle=0,width=7.0in]{vel_o_plot.eps}
%\caption{O star orbit }
%\label{Oorbit}
%\end{figure*}
%
%
%
%

\bsp \label{lastpage}

\end{document}